\documentclass[lineno]{jfm}

\usepackage{graphicx}
\usepackage{newtxtext}
\usepackage{newtxmath}
\usepackage{natbib}
\usepackage{hyperref}
\hypersetup{
	colorlinks = true,
	urlcolor   = blue,
	citecolor  = black,
}

\newcommand{\RomanNumeralCaps}[1]
\linenumbers

\usepackage{tikz}
\usetikzlibrary{positioning,shapes}
\usepackage{soul}

\usepackage{blindtext}

\shorttitle{TBL response to uniform changes of the pressure force contribution}
\shortauthor{T. R. Gungor}

\title{Turbulent boundary layer response to uniform changes of the pressure force contribution   }

\author{Taygun R. Gungor \aff{1,2},
	Ayse G. Gungor\aff{2},
	\and Yvan Maciel\aff{1}\corresp{\email{yvan.maciel@gmc.ulaval.ca}}
}

\affiliation{
	\aff{1} Department of Mechanical Engineering, Laval University, Quebec City, QC, G1V 0A6 Canada
	\aff{2}Faculty of Aeronautics and Astronautics, Istanbul Technical University, 34469 Maslak, Istanbul, Turkey
	
}

\begin{document}
	\maketitle
	
	\begin{abstract}

		We investigate a turbulent boundary layer (TBL) under uniform pressure force variations, focusing on understanding its response to local pressure force, local pressure force variation (local disequilibrating effect), and upstream history. The flow starts as a zero-pressure-gradient (ZPG) TBL, followed by a uniform increase in the ratio of pressure force to turbulent force in the outer region and concludes with a uniform decrease of the same magnitude. This last zone includes a subzone with a diminishing adverse-pressure-gradient (APG), followed by an increasing favorable-pressure-gradient (FPG). In both subzones, the impact remains the same: mean momentum gain and turbulence reduction. In the outer region, the mean flow responds to force balance changes with a considerable delay. The accumulated flow history leads to a FPG TBL at the domain’s end with a momentum defect comparable to APG TBLs. Below $y^+=10$, the mean flow responds almost instantaneously to pressure force changes. In the overlap layer, velocity profiles deviate from the conventional logarithmic law of ZPG TBL. Outer-layer turbulence decays more slowly than it initially increases, the latter persisting even after the pressure force begins to decrease. As a result of the slow turbulence decay, the FPG TBL at the domain's end exhibits unusually high outer turbulence levels. Near the wall, turbulence responds with a delay to pressure force changes, partly due to the influence of large-scale turbulence. All these significant cumulative effects of continuous pressure force variation indicate that parameters based solely on local variables cannot fully describe the physics of non-equilibrium TBLs.

	\end{abstract}
	
	\begin{keywords}
	\end{keywords}

	\section{Introduction}
	
	Most boundary layers found in industrial applications or in nature are complex because, besides being turbulent, they are affected by pressure gradient, wall curvature, wall roughness, heat transfer or a combination of these factors. As a result, one of the fundamental characteristics of these turbulent boundary layers (TBL) is that they are non-equilibrium flows. This paper focuses on non-equilibrium effects caused by the pressure gradient. Its premise is that in order to better understand and to better model turbulent boundary layers subjected to a pressure gradient, we have to understand, first and above all, how the pressure gradient unbalances the boundary layer.

	To avoid any confusion, it is important to define what is meant here by “flow equilibrium”. Flow equilibrium refers to the state where all forces acting on the fluid maintain the same balance as the flow develops \citep{rotta1953theory,clauser1956turbulent,townsend1956properties,maciel2006self}. The properties of an equilibrium flow, for instance its thickness, velocity, and forces’ magnitudes, can evolve in the streamwise direction but the balance of forces remains identical. The flow thus remains dynamically similar. It is important to note that in the context of turbulent flows, "equilibrium" can also have a different meaning; it is also employed to express particular types of turbulent energy balance, as discussed by \cite{spalart2015philosophies}. 
	
	In turbulent boundary layers, the situation is more complex than in laminar flows and turbulent free shear flows due to the presence of two dynamically distinct wall-normal layers, with independent length scales: the inner and outer layers. The balance of forces differs significantly between these layers. In the inner layer, viscous forces are important while mean inertia forces are small, whereas the opposite is observed in the outer layer. This means that the inner layer can reach a state of equilibrium or near equilibrium, for instance corresponding to the law of the wall, while the outer layer remains in a state of disequilibrium. When we refer to "near equilibrium," we mean that the balance of forces undergoes slow variation in the streamwise direction. The ZPG TBL serves as an example of a near-equilibrium TBL, where the force balance in the outer region experiences slow variations due to the increasing local Reynolds number. Unlike the sink-flow TBL, which is a single-layer flow \citep{townsend1956properties,rotta1962turbulent}, ZPG TBLs and FPG/APG “equilibrium” TBLs are anticipated to achieve exact equilibrium only in the theoretical scenario of infinite Reynolds number, owing to their two-layer nature.

	The concepts of self-similarity and equilibrium are interconnected. A flow or a region of a flow is self-similar if its statistical properties, or at least some of them, such as the mean velocity profile, depend solely on local flow variables (scales). If a flow region reaches a self-similar state, then it can only depend on its upstream history through its length and velocity scales. It has therefore become an equilibrium flow. A review of these ideas can be found in \cite{maciel2006self} and \cite{devenport2022equilibrium}. Equilibrium boundary layers are seldom encountered in the real world and are challenging to achieve experimentally and numerically. There exist only a few experimental and numerical realizations of equilibrium FPG and APG TBLs with varying degrees of success (see \cite{devenport2022equilibrium}). The interest of equilibrium TBLs lies in the fact that they represent theoretically simpler forms of TBLs, as they can be described by local parameters alone.

	The Rotta-Clauser pressure gradient parameter, $\beta_{RC}$, is a commonly used pressure gradient parameter in the literature. 
	
		\begin{equation}
		\beta_{RC} = \frac{\delta^*}{\rho u_\tau^2}\frac{dp_e}{dx}
		\label{betarc_eq}
	\end{equation}
	
	\noindent where $\rho$ is density, $\delta^*$ is the displacement thickness, $u_\tau$ is friction velocity and $p_e$ is the mean pressure at the edge of the boundary layer. It can be obtained from the momentum integral equation as the ratio of the total streamwise pressure force to the wall shear force acting on the whole boundary layer momentum defect. Therefore, it is not an inner or outer pressure gradient parameter but rather a global one that encompasses the entire boundary layer. Although this paper utilizes pressure gradient parameters specific to the inner and outer layers,  $\beta_{RC}$ will also prove useful in analyzing non-equilibrium effects in the current paper.

	In general PG TBLs, the overall local dynamic state of the inner region can be expressed with the pressure gradient parameter $\beta_i$ \citep{mellor1966effects}, which represents the ratio of pressure force and apparent turbulent force (gradient of Reynolds shear stress) 	
	\begin{equation}
		\beta_i = \Delta p^+ = \frac{\upnu}{\rho u_\tau^3}\frac{d p_w}{dx}
	\end{equation}

	\noindent where $\nu$ is viscosity and $p_w$ is the mean pressure at the wall. In the literature, $\beta_i$ is also denoted	$\Delta p^+$ or $p^+$ and considered as the local pressure gradient in wall units. Equilibrium in the near-wall region implies that $\beta_i$ remains constant. 
	
	Equilibrium in the outer region, including the overlap layer, is more difficult to achieve because of the inertia forces. For this reason, the various existing similarity analyses \citep{rotta1962turbulent,townsend1956properties,mellor1966equilibrium,castillo2001similarity,maciel2006self}  find necessary conditions for self-similarity, and thus equilibrium, expressed with outer layer parameters. The main (driving) condition is that the outer pressure gradient parameter, in any chosen form as long as it represents a ratio of forces, must remain constant. By using five non-equilibrium and one equilibrium APG TBL databases, \cite{maciel2018outer} showed that the ratio of forces in the outer layer can be accurately followed with three parameters when these parameters are expressed with Zagarola-Smits scales as the outer scales: $U_o=U_{ZS}=(\delta^*/\delta)U_e$ and $L_o =\delta$, where $\delta$ is the boundary layer thickness and $U_e$ is the velocity at the edge of the boundary layer. Such a correspondence was not achieved with the friction velocity or the outer pressure-gradient velocity scale, $U_{po}^2=(\delta^*/\rho)(dp_e/dx$) whatever the length scales chosen.  The outer pressure gradient parameter is therefore expressed with Zagarola-Smits scales in the present study: 
	
	\begin{equation}
		\beta_{ZS}=\frac{\delta}{\rho U_{ZS}^2}\frac{dp_e}{dx}
	\end{equation}
	
	\noindent and it precisely follows the ratio of pressure force to turbulent force in the outer region of the six APG TBLs in the study of \cite{maciel2018outer} and in the APG TBL region of the current flow. The present study suggests that the correspondence is not as accurate for FPG TBLs, at least for the present one which is recuperating from strong downstream APG effects, but $\beta_{ZS}$ still follows the variation of the force ratio in the FPG region of the flow.

	From the above discussion, it becomes clear that a TBL which is in a non-equilibrium state in the outer region because of the pressure force is basically a TBL for which the pressure gradient parameter $\beta_{ZS}$ varies.  It is important to stress that it is not the pressure gradient in itself that causes a non-equilibrium state, but rather the streamwise evolution of the pressure gradient parameter. In other words, $d\beta_{ZS}/dx$ is a measure of the local disequilibrating effect of the pressure force whereas $\beta_{ZS}$ is a measure of the local direct effect of the pressure force. This study illustrates this point further by showing that the sign of the pressure gradient does not always reflect the type of disequilibrating effects. It will be seen that, as counterintuitive as it may appear, a positive pressure gradient ($dp/dx>0$; APG) can lead to the boundary layer filling up with momentum if $d\beta_{ZS}/dx<0$. We will call the disequilibrating effects of $d\beta_{ZS}/dx<0$ as momentum-gaining conditions since the qualifier favorable refers in the literature to $dp/dx<0$.

	Besides not fully understanding such local pressure gradient effects on the evolution of a TBL, we are unable to make a clear distinction between them and Reynolds number and upstream history effects. In the case of mild APG effects, \cite{vila2020separating}, \cite{pozuelo2022adverse}, and \cite{deshpande2023reynolds} provide insights on the differences between APG and Reynolds number effects. If the imposed pressure gradient effect is strong (large $\beta_{ZS}$ and/or $d\beta_{ZS}/dx$) and the Reynolds number is high, Reynolds number effects can probably be neglected. The question of upstream history effects is more complex since they cannot be neglected and they can hardly be represented by a local parameter. Upstream history effects are, by definition, cumulative effects that can only be taken into account with the integration of the governing equations. They have been studied for different types of non-equilibrium PG TBLs, including cases where the transition from APG to FPG, or vice-versa, is considered. Three recent studies have focused on comparing APG TBLs with different upstream histories at locations where both $\beta_{RC}$ and the friction Reynolds number $Re_{\tau}$ match \citep{bobke2017history,vinuesa2017revisiting,tanarro2020effect}. The findings revealed that upstream non-equilibrium APG conditions affect the outer region more significantly than the inner region. The latter is found to be more dependent on the local value of $\beta_{RC}$. However, these studies assume that upstream history effects are the same for both the inner and outer regions by using only $\beta_{RC}$ as the pressure gradient parameter, but in reality, they differ. 
	
	Although the boundary layer does not respond in the same manner to continuous gradual changes in the force balance and to rapid or step changes of flow conditions, the latter cases can nonetheless help us understand the response of the boundary layer to non-equilibrium conditions. From the equations of motion, it can be deduced that the mean shear rate $\partial U/\partial y$ in the outer region cannot respond immediately to a step change in the pressure gradient since viscous diffusion is negligible there (Smits and Wood, 1985). Consequently, turbulence in the outer layer cannot also respond immediately to a step change in the pressure gradient. Based on scaling arguments, \cite{devenport2022equilibrium} suggest that it takes tens of boundary layer thicknesses for both the mean flow and turbulence to adjust to pressure gradient changes. This is consistent with the boundary layer recovery distances observed by Volino (2020) downstream of a FPG to ZPG transition.

	The extended recovery in the outer region is also observed in channel flows with a step change from rib- or cube-roughened surface to a smooth wall \citep{ismail2018simulations} and in pipe flows subjected to various strong localized perturbations \citep{ding2021relaxation, van2020turbulent, ding2021perspective}. In their studies on pipe flows, Smit’s research group found that turbulence recovery in the far-field was oscillatory, resembling a second-order dynamical system, overshooting the eventual fully-developed condition. This oscillatory behavior is due to the asynchronous recovery between the mean velocity and the Reynolds shear stress.

	In contrast to the situation in the outer layer, turbulence in the inner layer adjusts itself very quickly. For instance, in experiments with rods placed inside ZPG TBLs \citep{clauser1956turbulent,marusic2015evolution}, the inner layer recovers after a few boundary layer thicknesses, while the outer layer takes tens or even hundreds of boundary layer thicknesses to recover. A similar situation occurs 	in the TBL recovery from separation due to a step, where the outer region is strongly energized by the pertubation \citep{vaquero2022outer}. Cases of a very fast transition from FPG to APG \citep{tsuji1976turbulent, parthasarathy2023family} also show that the near-wall region adapts more rapidly than the outer region. In such cases, a new internal layer grows inside the APG TBL, starting from the wall, in reaction to the change, while the upper part of the TBL responds much more slowly. \cite{parthasarathy2023family} found that this internal layer establishes its own inner and outer regions. The triggering of an internal layer is more commonly found in TBLs with abrupt changes of wall conditions --- surface curvature, roughness, wall heat flux (Smits and Wood 1985). In the case of rough-wall TBLs with step changes in roughness, it has been shown that the outer-layer recovery is nonlinear and depends on the sign of the step change (Smits and Wood 1985). The return to an equilibrium state in the outer region is more rapid when turbulence is augmented (smooth to rough step change) than when turbulence is reduced (rough to smooth step change). Turbulence appears to take more time to decay than to build up in the outer region of TBLs.

	The response of turbulence in the outer region is known to lag even in cases of continuous changes of the pressure force. For example, \cite{devenport2022equilibrium} remark that at the streamwise location where $\beta_{RC}\approx 2$, the Reynolds shear stresses in the cases of an increasing pressure force effect ($d\beta_{RC}/dx>0$) of \cite{nagano1993effects} and \cite{spalart1993experimental} are about half of those of the equilibrium case of \cite{bobke2017history}.

	To better understand the direct local pressure force effect, the local disequilibrating effect of the pressure force and the upstream history effect, we have designed a non-equilibrium flow with an almost linear increase followed by an almost linear decrease of the pressure gradient parameter $\beta_{ZS}$, as seen in figure \ref{fig:chars}$(a)$. The local pressure gradient effects start therefore as momentum-losing conditions, of constant strength, followed by momentum-gaining conditions of the same strength. The momentum-gaining conditions ($d\beta_{ZS}/dx<0$) are maintained on a long length to also include a zone of FPG ($dp/dx<0$) at the end of the domain. Note that the middle region of the flow is a zone of APG ($dp/dx>0$) that is momentum-gaining for the flow ($d\beta_{ZS}/dx<0$). In other words, the boundary layer is filling up with momentum even if the external flow is still decelerating. Although this may appear counterintuitive as stated above, it is consistent with an analysis of the mean momentum equation, boundary layer computations of the current flow, and the results presented in the paper. This is a unique feature that has not been specifically studied, although it is also encountered in the TBLs alternating from mild APG to mild FPG of \cite{tsuji1976turbulent} or \cite{fritsch2022experimental}.

	The current TBL is experiencing moderate APG disequilibrium (indicated by a moderate value of $d\beta_{ZS}/dx>0$). However, due to the long length of the imposed pressure force increase, it attains a significant velocity defect ($H=2.87$). This places it between quasi-equilibrium APG TBLs with large velocity defects \citep{kitsios2017direct,skaare1994turbulent} and our recent APG TBL, which is in a state of strong disequilibrium \citep{maciel2018outer,gungor2022energy}. This feature, combined with the previous one, makes it an excellent intermediate test case to determine what distinguishes equilibrium and disequilibrium APG TBLs in strong pressure gradient conditions. Likewise, the slow deceleration of the TBL in a long domain helps us determine under what conditions the classical law of the wall starts to lose its validity.

	Because of the switch from momentum-losing to momentum-gaining pressure conditions, it is possible to study the delayed response of the mean flow and turbulence in both the inner and outer layers. Additionally, the final segment of the domain represents an FPG TBL at high Reynolds number with a distinctly unique flow history.
	
	The present work aims to better understand and distinguish the three different effects of the pressure force---local direct, local disequilibrating and upstream variation---on turbulent boundary layers. 

	\section{The database}

	\subsection{The numerical details}

	\begin{figure}
		\centering                     
		\begin{tikzpicture}   
			\hspace{-0.5cm}                     
			\node(a){ \includegraphics[scale=0.7]{ 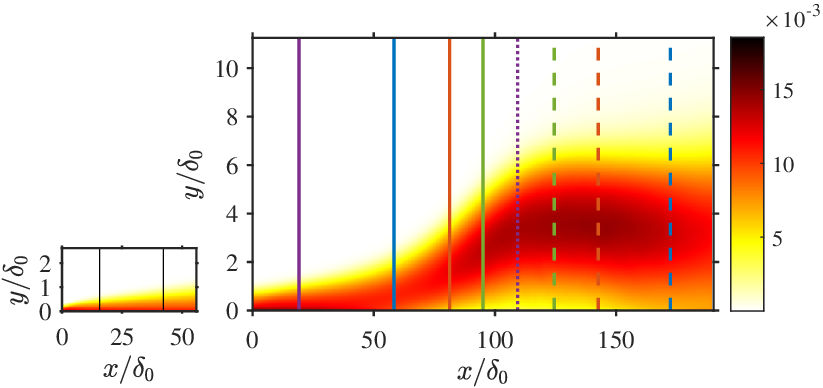}};
			\node at  (-3.7,-0.4) [overlay, remember picture] {$\pi_r$};	
			\node at  (-3,-0.4) [overlay, remember picture] {$\pi_t$};		\end{tikzpicture}

		\vspace{-.1cm}

		\caption{The spatial evolution of the turbulent kinetic energy, normalized with $U_{e,0}$ and $\delta_0$, in the auxiliary (left) and main (right) domain as a function of $x/\delta_0$ and $y/\delta_0$ where $\delta_0$ is the boundary layer thickness at the inlet of the main domain. $\pi_r$ and $\pi_t$ indicate the location of the recycling and transfer planes. The vertical lines on the main domain indicate the streamwise stations selected for the detailed analysis. 	The $x$-axis is not scaled proportionally to the $y$-axis. Only the lower half of the domain is shown for both domains.}
		\label{sche}
	\end{figure}

	The novel flow case is generated using a DNS code that employs a fractional step method to solve the three-dimensional incompressible Navier-Stokes equations with primitive variable formulation within a three-dimensional rectangular volume. The methodology for this DNS code is based on \cite{kim1985application}. In terms of computational specifics, the grid is structured and staggered. Spatial discretization entails a fourth-order compact finite difference scheme for convective and viscous terms, as well as a standard second-order discretization for the pressure term in both the streamwise and wall-normal directions. The spanwise direction uses spectral expansion, de-aliased via the 2/3 rule, for discretization \citep{lele1992compact}. Temporal advancement is accomplished through a semi-implicit three-step Runge-Kutta method. For more comprehensive insights into the DNS code and further particulars, refer to \cite{simens2009high}, \cite{borrell2013code}, and \cite{sillero2014high}.
	
	The DNS computational setup consists of two simultaneous simulation domains, depicted in figure \ref{sche} and elaborated further in \cite{borrell2013code}. The auxiliary DNS, possessing coarser resolution, aims to deliver realistic turbulent inflow data for the primary DNS. It involves a ZPG TBL, where inlet conditions are derived from a recycling plane within the domain. Data from this recycling plane ($\pi_r$ in figure \ref{sche}) are rescaled using a modified version of method of \cite{lund1998generation} \citep{simens2009high}. The positioning of the recycling plane with respect to the inlet is determined to ensure sufficient distance for flow decorrelation from inlet effects \citep{simens2009high}. A transfer plane ($\pi_t$ in figure \ref{sche}) from the auxiliary simulation is designated as the source of temporal data for inflow conditions in the main simulation. For the transfer plane, the data from the auxiliary domain are extrapolated in the freestream because the wall-normal height of both domains are different from each other. The recycling and transfer planes are located at approximately $x=46\delta_{0,a}$ and $124\delta_{0,a}$, where $\delta_{0,a}$ is $\delta$ at the inlet of the auxiliary domain. The adoption of two domains is needed to provide inflow conditions for the main DNS at a lower computational cost.

	Regarding the other boundary conditions, the lower surface represents a flat plate with a no-slip (zero velocity) and impermeability condition. As for the upper boundary, wall-normal velocity is applied via suction and blowing, generating favorable/adverse pressure gradients. Additionally, the free-slip conditions are enforced for the wall-parallel components. At the exit of the domain, convective boundary conditions are employed. In addition, the auxiliary domain was initialized with a flow realization of a ZPG TBL.
	
	The computational domains of the main and auxiliary cases are rectangular volumes. The domain length and grid properties are given in table \ref{domain} for both cases. Here, the average boundary layer thickness ($\delta_{av}$) is calculated within the useful range of the main domain. The useful range is 36$\delta_{av}$ long. For domain sizes, the wall-normal length of the domain is chosen as approximately $3$ times $\delta$ at the exit and spanwise length is chosen based on two-point correlations. The Reynolds number based on momentum thickness ($Re_\theta$) of the main case ranges from $1941$ to $12970$. 
	
	Spatial resolutions, $\Delta x^+$ and $\Delta z^+$, range from $1.56-9.84$, and $1.18-7.37$, respectively, with the coarsest resolutions near the inlet and finest resolutions where the flow exhibits a significant velocity defect. Furthermore, $\Delta y^+$ at the wall does not exceed $0.49$, and the maximum $\Delta y^+$ across the boundary layer remains below $11.06$. The time step size, $\Delta t^+$, is maintained below $0.17$ throughout the domain.

	The data for the statistics were collected by averaging over time and the spanwise direction. The duration of data collection was $11.5$ and $7.7$ flow-through times based on the edge velocity, $U_e$, at the inlet and exit, respectively. The eddy-turnover time, based on $u_\tau$, is highest at the inlet, $85.9$. It decreases to $2.2$ at $x/\delta_{av} \approx 20$, which is very close to SP5. However, in TBLs with a large-velocity defect, $u_\tau$ is no longer a characteristic scale for large-scale structures. Therefore, we also computed the eddy-turnover time using $U_{ZS}$, which gives an eddy-turnover time of $323$ at the inlet and a minimum value of $21$. The simulations were performed on the Marconi KNL and Marconi100 clusters at Cineca, as well as on the Niagara cluster of the Digital Alliance of Canada. The cost of the simulation for collecting statistics is approximately 10 million core-hours on Niagara of Digital Alliance of Canada.

	\begin{table}
		\centering
		\caption{Domain properties of the main and auxiliary DNS. $\delta_{av}$ is the average thickness of the boundary layer of the main domain in the useful range and $\delta_e$ is the thickness of the boundary layer at the exit of each domain. }		\begin{tabular}{cccccccc} 
			&  $(L_x, L_y, L_z)/\delta_{av}$ & $(L_x, L_y, L_z)/\delta_{e}$ & $N_x, N_y, N_z$ & $Re_\theta$ &  $Re_\tau$ & \\ \hline
			Auxiliary \\ domain  & $12.0, 1.2, 6.3$ &  $46.7,4.2,16.5$ & $2161,316,2700$ & 620-2380   & 260-810\\ \hline
			Main \\ domain &  $40.2,4.7,6.3$ & $26.1,3,4.1$ & $12801, 770, 2700$ & $1941$-$12970$ & $654-2489$  \\ \hline
			& & & & & & 
		\end{tabular}
		
		\label{domain}
	\end{table}

	\section{Results }
	
	Before discussing the results, the edge of the boundary layer needs to be clearly defined. The definition of the boundary layer edge and edge velocity in the case of pressure gradient TBLs and TBLs over complex geometries is challenging due to variations in freestream velocity. \cite{griffin2021general} provide a review of eight methods of computing the boundary layer thickness in such cases. We have tested several methods, and a summary of the main results are presented in Appendix A. The method based on the local reconstruction of the inviscid-flow velocity profile by \cite{griffin2021general} is the most theoretically rigorous and empirically consistent. Unfortunately, we cannot use it in the present study because it requires mean static pressure data, which we lack for the near-equilibrium TBL database of \cite{kitsios2017direct} used for comparison. Instead, we chose the recent method of \cite{wei2023outer}, because it yielded the most consistent results across all tested databases, although, like most methods, it lacks a theoretical basis. In this method, the boundary layer thickness, represented as $\delta$, is defined as the location near the freestream where the Reynolds shear stress reaches 5\% of its maximum. The edge velocity, denoted as $U_e$, is defined as the mean streamwise velocity at that location.

	\begin{figure}
		\begin{tikzpicture}   
			\centering                     
			\node(a){ \includegraphics[scale=0.6]{ 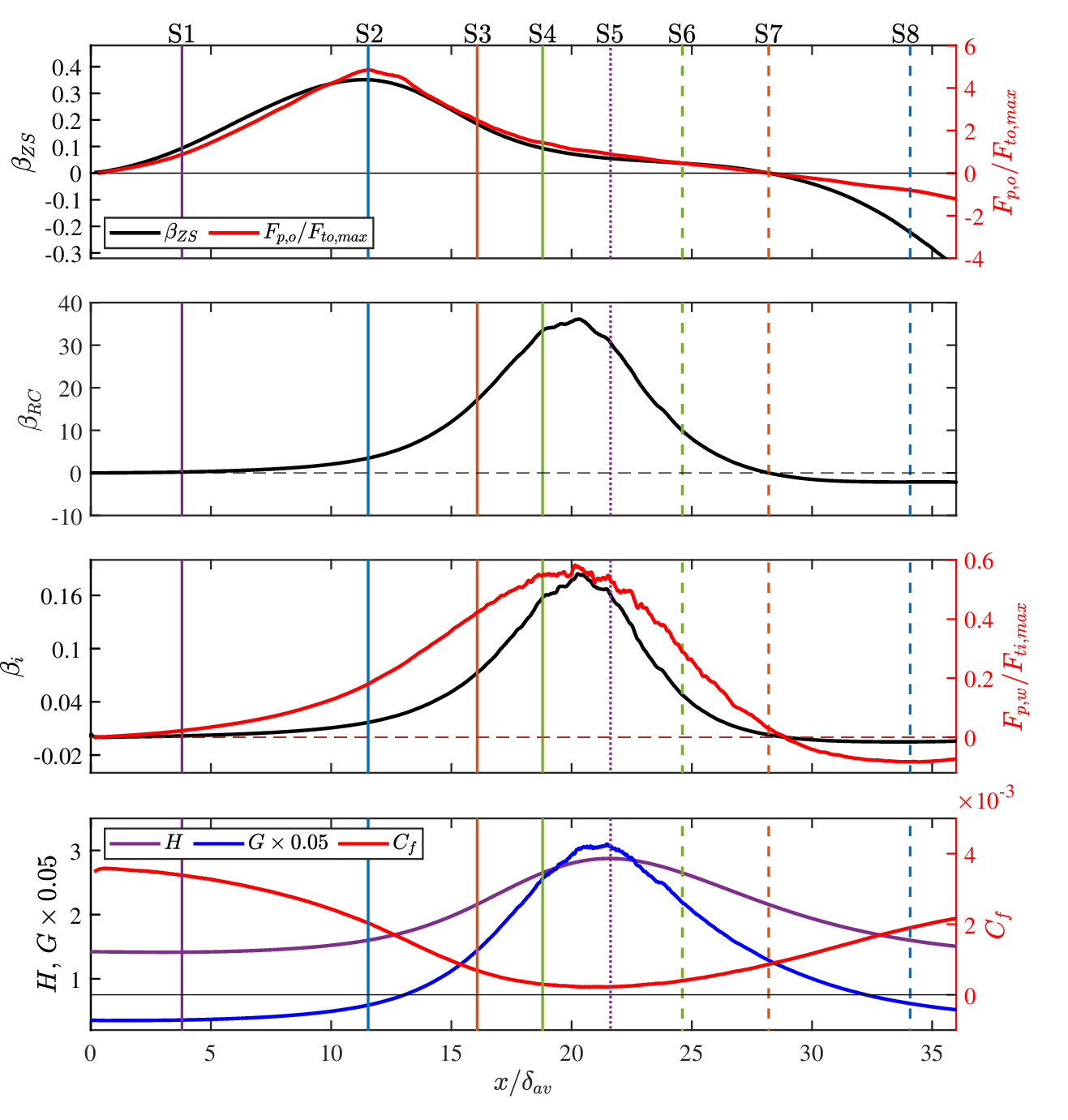}};
			\node at  (-6.5,3) [overlay, remember picture] {($b$)};		
			\node at  (-6.5,-0.2) [overlay, remember picture] {($c$)};				
			\node at  (-6.5,-3.4) [overlay, remember picture] {($d$)};					\end{tikzpicture}
		
		\vspace*{-0.5cm}
		\caption{Streamwise development of the main parameters of the APG TBL. The vertical lines indicate the streamwise stations selected for the detailed analysis.}
		\label{fig:chars}
	\end{figure}
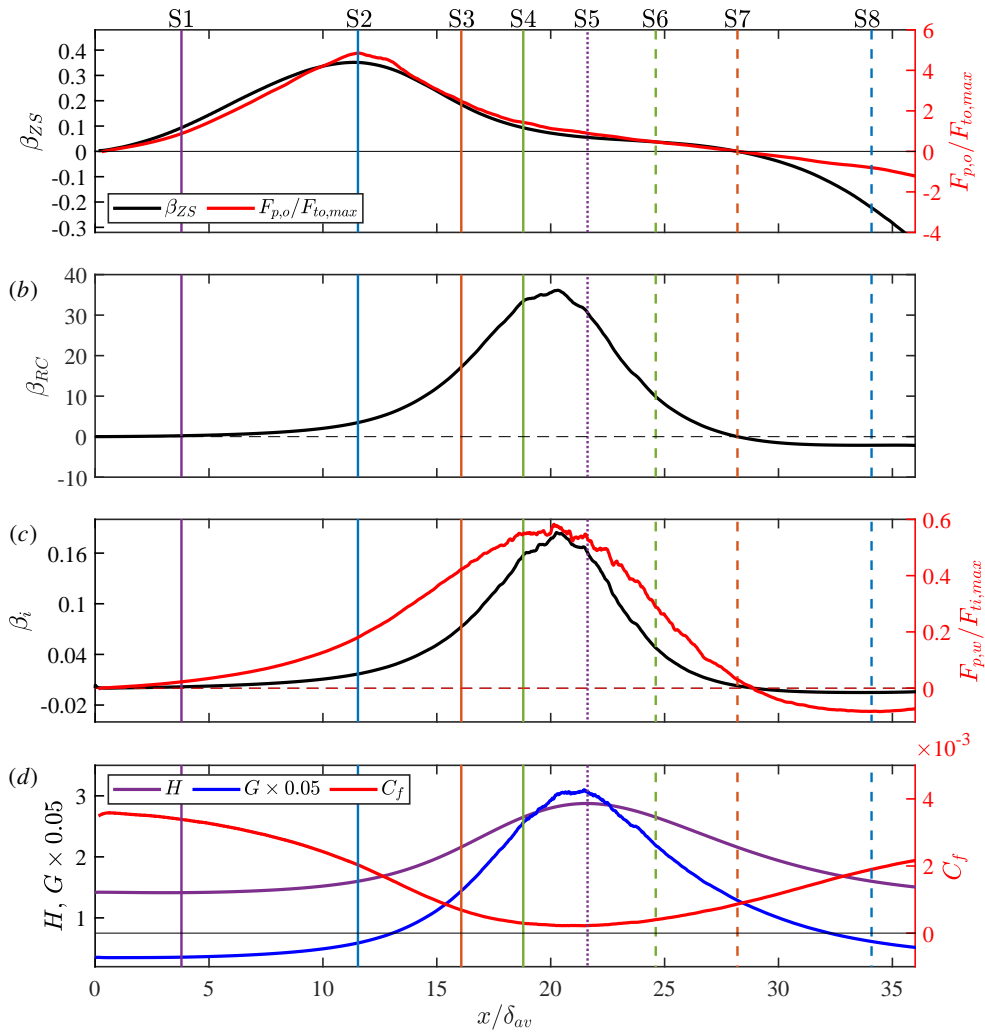

	\subsection{The mean flow and force balance }

	To first understand the local and upstream effects of the pressure force on the mean flow, we investigate global flow parameters, mean momentum equation budgets, and mean velocity profiles separately for the outer and inner layers. The streamwise mean momentum equation where each term can be viewed as a force acting on the flow is given in equation \ref{mombud}.

	\begin{equation}
		0 = 
		\bigg (-U \frac{\partial U}{\partial x}-V\frac{\partial U}{\partial y} \bigg )-
		\frac{1}{\rho}\frac{dp_e}{dx}-
		\frac{\partial 	\langle u v\rangle}{\partial y}+
		\upnu\frac{\partial^2 U}{\partial y^2}
		\label{mombud}
	\end{equation}
	
	\noindent where $U$ and $V$ are the mean velocities in the streamwise and wall-normal directions, $u$ and $v$ are the fluctuation velocities, $\upnu$ is the viscosity and $\langle.\rangle$ indicates ensemble averaging. The terms on the right-hand side of the equation represent the inertia force, the pressure force, the turbulent force (the gradients of the normal stresses are not considered because they are negligible) and the viscous force, respectively.

	\subsubsection{Outer layer}

	The global parameters of the flow and the ratio of forces in the inner and outer layers are presented in figure \ref{fig:chars} as a function of $x/\delta_{av}$. Figure \ref{fig:chars}$(a)$ presents the outer-layer pressure gradient parameter $\beta_{ZS}$ and the ratio of pressure force at the edge of the boundary layer ($F_{p,o}$) to the maximum of turbulent force in the outer layer ($F_{to,max}$), where the outer layer is defined as the region above $0.1\delta$. \cite{maciel2018outer} demonstrated that $\beta_{ZS}$ follows the ratio of pressure to turbulence force in the outer region using six APG TBLs. This is confirmed in the present flow as $\beta_{ZS}$ and the ratio of forces follow each other very closely in the region where the pressure gradient is positive (adverse). However, $\beta_{ZS}$ and the force balance do not correspond in the FPG region, although the trend is still the same for both. This could be due, at least partly, to the delay in turbulence response to the flow changes. As it is presented in Section 3.2, Reynolds shear stresses remain very high in the outer region of the whole momentum-gaining zone, even higher than downstream. Their expected decay only starts at $x/\delta_{av}\approx28$. This could explain why the absolute value of the ratio of pressure force to turbulence force is low.

	As it was already presented in the introduction, in the first part of the domain, until approximately $x/\delta_{av}=12$, $\beta_{ZS}$ increases at a fairly constant rate as it was aimed for in the design of this flow. Afterwards, it decreases at a similar absolute rate until the end of the domain. Such a streamwise evolution of $\beta_{ZS}$ implies a non-equilibrium boundary layer with a constant increase in importance of the pressure force with respect to the other forces followed by a constant decrease as it can be seen with $F_{p,o}/F_{to,max}$. Until  approximately $x/\delta_{av}\approx21.5$ where the shape factor reaches its maximum (figure \ref{fig:chars}($d$)), the flow is similar to some strong non-equilibrium APG TBL cases in the literature \citep{gungor2016scaling,gungor2022energy,hosseini2016direct, hickel2008implicit}. In this region, the boundary layer constantly loses momentum ($H$ increases) because the pressure force's importance increases upstream for a long extent, as reflected by $F_{p,o}/F_{to,max}$ and $\beta_{zs}$. Afterwards, the flow is still under the effect of an adverse pressure gradient until $x/\delta_{av}=28$. However, the flow gains momentum in this region because the pressure force's importance decreases ($\beta_{ZS}$ decreases), which is a particular behaviour not studied in the past, as discussed in the introduction. The rest of the domain is an FPG TBL where the pressure gradient is maintained with a negative $d\beta_{ZS}/dx$ similar to that upstream.

\begin{figure}
		\centering
		\includegraphics[scale=0.8]{ 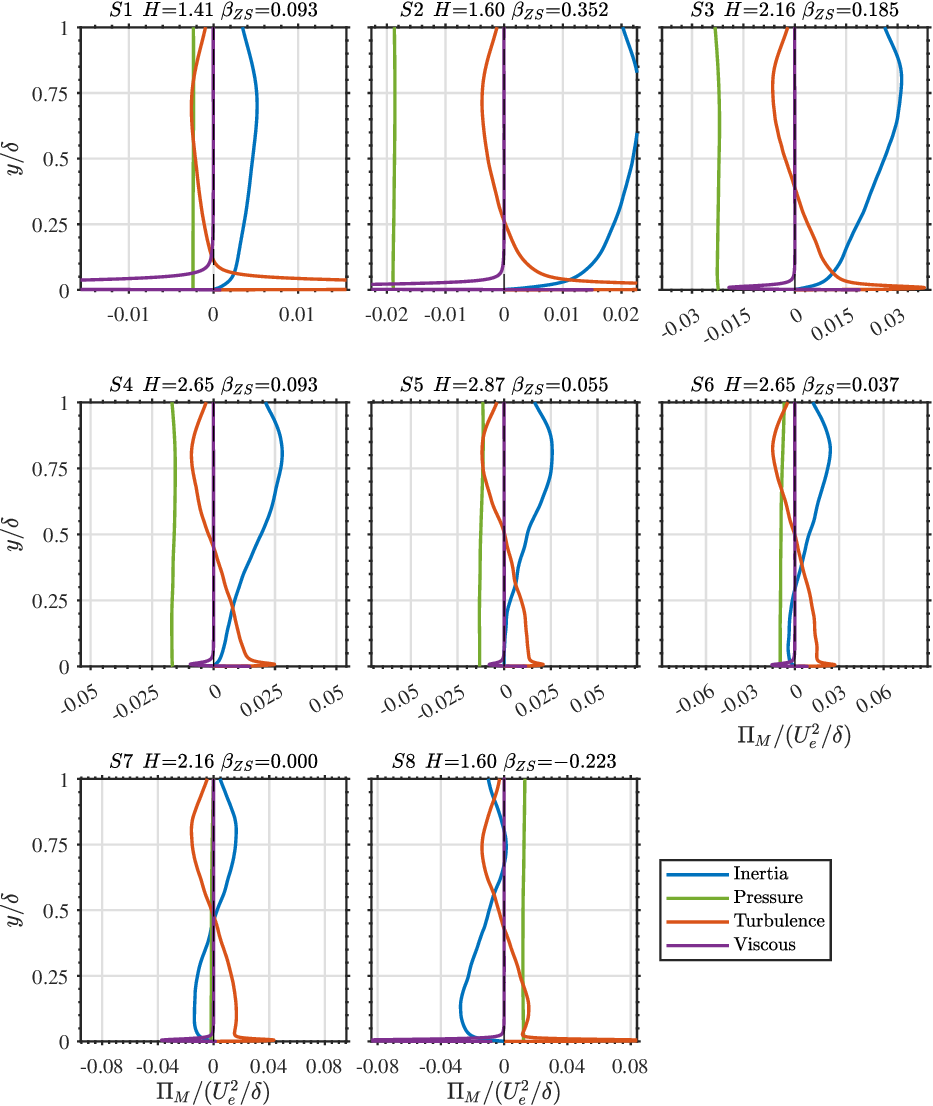}
		\caption{The mean momentum budget profiles of the eight streamwise positions as a functions of $y/\delta$. The budget terms are normalized with $U_{e}$ and $\delta$. The $x$-axis ranges are chosen so that the reference turbulent force value $F_{to,max}$ is approximately at the same location in all sub-figures. }
		\label{mean_mom_outer}
	\end{figure}

\begin{figure}
		\centering
		\hspace*{-1.5cm}
		\begin{tikzpicture}   
			\node(a){ \includegraphics[scale=0.6]{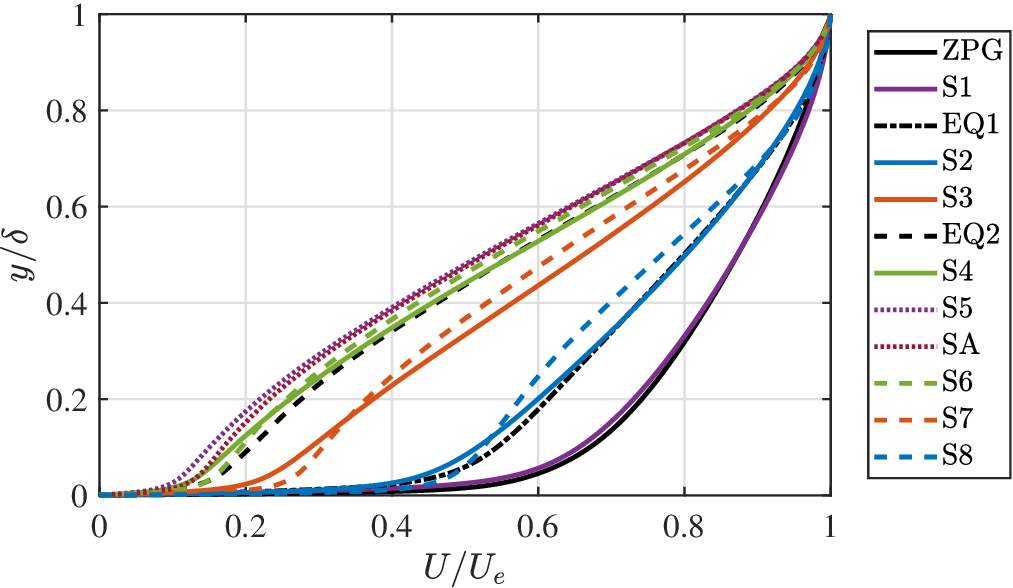}};
		\end{tikzpicture}  
		\caption{The mean velocity profiles for nine streamwise positions of the present case along with the ZPG TBL and near-equilibrium cases EQ1 and EQ2 as a function of $y/\delta$. The profiles are normalized with $U_e$.   }
		\label{fig:mean_vel}
	\end{figure}

	While we will present the remaining parameters in figure \ref{fig:chars} later, for now, we focus on discussing the mean momentum budgets to provide a more comprehensive description of the streamwise evolution of the force balance. Figure \ref{mean_mom_outer} presents the outer-scaled mean momentum budgets for various streamwise positions. These positions are identified with vertical lines in figure \ref{fig:chars}. Details of these positions, including the acceleration parameter ($K=(\nu U^2)/(dU_e/dx)$), are given in table \ref{table1} and the corresponding mean velocity profiles are shown in figure \ref{fig:mean_vel}. In figure \ref{mean_mom_outer}, the $x$-axis ranges have been selected to ensure that the reference turbulent force value ($F_{to,max}$) is approximately located at the same position in all sub-figures.

	The momentum budgets reveal the non-equilibrium nature of the flow since the balance of forces changes from one station to another. The budgets of the first two stations (S1 and S2) show explicitly that the pressure force increases in importance with respect to the turbulent force in the outer region. Further downstream, the pressure force's relative magnitude with respect to the other forces begins to decrease as can be seen with the momentum budgets starting from S3. This decrease continues until the pressure force becomes zero at S7. Afterwards, the pressure gradient becomes negative (FPG zone) until the end of the domain.

	In the momentum budgets of S1 to S6, the inertia term indicates that the fluid is losing momentum in the streamwise direction ($dU/dx<0$), except near the wall for S6. This is confirmed by the mean velocity profiles in figure \ref{fig:mean_vel}, but it is important to note that these are normalized profiles $U/U_e$ as a function of $y/\delta$ that are representative of $U$ variation in approximate streamline directions, not $x$. This explains why the fluid has gained momentum from S5 to S6 in the normalized representation of figure \ref{fig:mean_vel} although figure \ref{mean_mom_outer} indicates that $dU/dx$ is still negative at S6 in most of the outer region.

	Indeed, while it was constantly losing momentum upstream, the fluid starts to gain momentum around S5 because of the decrease of the pressure force's relative magnitude ($d\beta_{ZS}/dx<0$) that began slightly after S2. The fact that momentum gain only starts at S5, and not right after S2, shows that the mean flow responds with a delay to changes in the force balance because of inertia. As explained in the introduction, it is not the sign of the pressure gradient that indicates momentum gain or loss (disequilibrium), but the streamwise evolution of the relative magnitude of the pressure force, reflected by $d\beta_{ZS}/dx$ in the outer layer and $d\beta_{i}/dx$ in the inner layer. The present flow was specifically designed to produce a zone, from S5 to S7, where the fluid gains momentum in the boundary layer (in the normalized representation $U/U_e$ as a function of $y/\delta$ of figure \ref{fig:mean_vel}) even if the pressure gradient is still positive (APG).

	\begin{table}
		\begin{center}
			\def~{\hphantom{0}}
			\begin{tabular*}{\textwidth}{ @{\extracolsep{\fill}} rrrrrrrrrrrr}
				Position & $x/\delta_{av}$ & $H$ & $C_f$ & $Re_\tau$ & $Re_\theta$ & $\beta_{ZS}$ & $\beta_i$& $\beta_{RC}$ & $K\times$ $10^{-6}$ \\
				S1 & 3.8  & 1.41 & 0.0034 & 809 & 2486  & 0.093  & 0.0016  &  0.22 & 0.18  \\
				S2 & 11.5 & 1.60 & 0.0020 & 1012 & 4931  & 0.352 & 0.0168  &  3.47 & 0.53  \\
				S3 & 16.1 & 2.16 & 0.0007 & 865  & 7961  & 0.185 & 0.0729  & 17.21 &  0.54  \\
				S4 & 18.8 & 2.65 & 0.0003 & 702  & 9635  & 0.093 & 0.1575  &  33.49 & 0.29 \\
				S5 & 21.6 & 2.87 & 0.0002 & 709  & 10952 & 0.055 & 0.1607  & 30.59 & 0.18  \\
				S6 & 24.6 & 2.65 & 0.0004 & 1006 & 12099 & 0.037 & 0.0476 & 9.83 & 0.11  \\
				S7 & 28.2 & 2.16 & 0.0009 & 1489 & 12827 & 0.000 & 0.0028  &  0.00 & -0.01 \\ 
				S8 & 34.1 & 1.60 & 0.0019 & 2181 & 12278 &-0.223 & -0.0051 & -2.11 & -0.12 \\ 
				ZPG & N/A & 1.38  & 0.0030 & 1334 &	3904  & 0.000 & 0.0000 & 0.00  & -  & \\
				EQ1 & N/A & 1.57  & 0.0025 & 811  & 3358  &	0.104 & 0.0056 & 1.04  & - &\\
				EQ2 & N/A & 2.57  & 0.0004 & 698  & 8584  & 0.090 & 0.1058 & 34.24 & - & 
			\end{tabular*}
			\caption{The main parameters of selected streamwise positions in the present case along with the ZPG TBL of \cite{sillero2013one} and the near-equilibrium APG TBL cases of \cite{kitsios2017direct}. }
			\label{table1}
		\end{center}
	\end{table}

	The momentum changes observed in figure \ref{fig:mean_vel} can also be followed with the distribution of the shape factor, $H$, as presented in figure \ref{fig:chars}$(d)$, with increasing $H$ indicating momentum loss and vice versa. The velocity defect is small at S1 ($H$=$1.41$), increases to become large and maximal at S5 ($H$=$2.87$) and then monotonously decreases downstream. The shapes of the profiles in the momentum-losing (streamwise positions S2 to S4) and -gaining zones (S6 to S8) are different from each other at matching or very close shape factors as illustrated in figure \ref{fig:mean_vel}. This happens due to the different upstream flow history in the two zones. The main difference is that the profiles in the momentum-gaining region are fuller in the inner layer due to the faster response of the flow in that layer.

	To examine how the momentum gain starts in the boundary layer, we consider two streamwise positions, S5 and the position denoted as SA, where the shape factor is 2.78, located just slightly downstream of S5. As it can be seen from figure \ref{fig:mean_vel}, the inner layer gains momentum at SA but the defect in the outer layer remains almost the same. The inner layer therefore starts gaining momentum before the outer layer even if the momentum-gaining effect of the pressure force initiates earlier in the outer layer ($d\beta_{zs}/dx$ becomes negative before $d\beta_i/dx$). This implies that the inner layer reacts to the pressure gradient much more rapidly than the outer layer. This is consistent with the fact that, in the inner layer, inertia forces are small and turbulence adjusts more quickly.

As mentioned in the introduction, $\beta_{RC}$ is not an outer pressure gradient parameter, but a global pressure gradient parameter for the whole boundary layer. It is presented in figure \ref{fig:chars}($b$) and a comparison of $F_{p,o}/F_{to,max}$ and $\beta_{RC}$ shows that $\beta_{RC}$ indeed is not an outer region pressure gradient parameter for large defect TBLs. It follows in fact more closely, but not exactly, the ratio of pressure to turbulent forces in the inner region given by the inner-layer pressure gradient parameter, $\beta_i$ (figure \ref{fig:chars}$(c))$.

	\begin{figure}
		\centering 
		\includegraphics[scale=0.6]{ 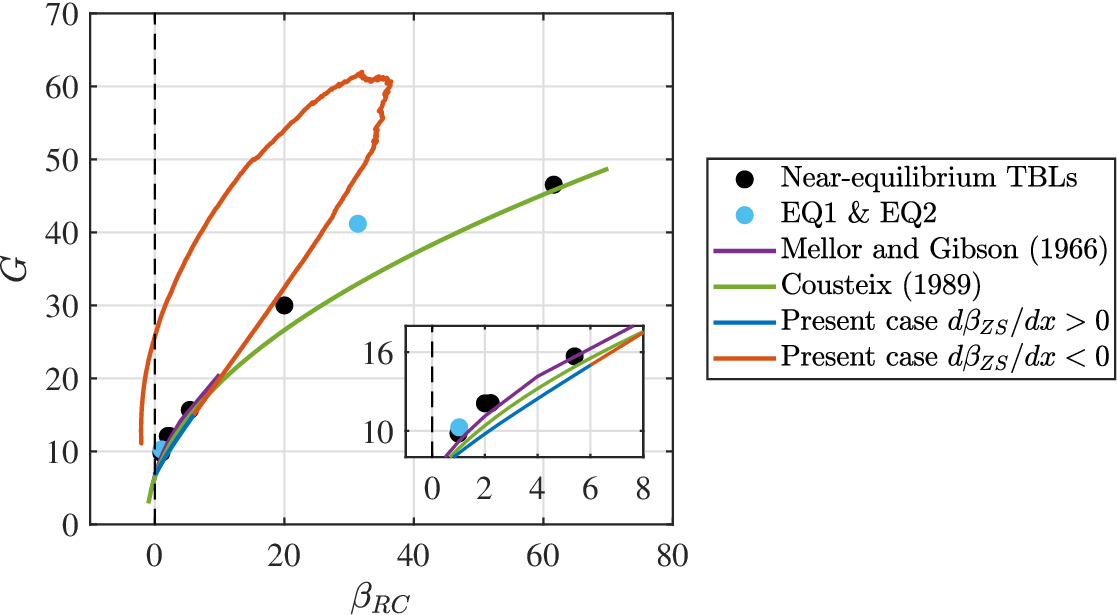}  
		
		\caption{The defect shape factor as a function of $\beta_{RC}$ for the present DNS (blue and orange are for the regions where $d\beta_{ZS}/dx>0$ and $d\beta_{ZS}/dx<0$, respectively), near-equilibrium cases from the literature \citep{kitsios2017direct,bobke2017history,vila2020separating,skaare1994turbulent,east1980investigation,bradshaw1967turbulence}, computations of \cite{mellor1966equilibrium}, and correlation of \cite{cousteix1989turbulence}. }
		\label{ggg_beta}
	\end{figure}

	To enhance our understanding of how the present flow responds to disequilibrium conditions, figure \ref{fig:mean_vel} illustrates mean velocity profiles from the two near-equilibrium APG TBLs of \cite{kitsios2017direct}. These near-equilibrium flows correspond to cases of small velocity defect ($H=1.57$), denoted herein as EQ1, and large velocity defect ($H=2.57$), denoted as EQ2. The main parameters of these flows are given in table \ref{table1}. Figure \ref{fig:mean_vel} illustrates that S2 and EQ1, sharing similar $H$ values, exhibit comparable velocity, but the near-equilibrium profile is fuller in the inner region. However, the near-equilibrium TBL attains this state with a globally smaller local pressure force effect (smaller $\beta_{RC}$) and a smaller effect in the outer region (smaller $\beta_{ZS}$). This suggests that the current non-equilibrium flow does not attain the anticipated equilibrium state corresponding to a given local pressure force effect ($\beta_{ZS}=0.352$ and $\beta_i=0.0168$). In other words, if the flow were to respond immediately to the pressure force, S2 would exhibit a larger velocity defect than EQ1, given that $\beta_{ZS}=0.352$ at S2 compared to $\beta_{ZS}=0.104$ at EQ1.

	\cite{mellor1966equilibrium} showed that for equilibrium boundary layers (at infinite Reynolds number) there is a unique relationship between the Rotta-Clauser defect shape factor $G$ and $\beta_{RC}$, where $G=(U_e/u_\tau)(1-1/H)$. The relationship was obtained numerically from their similarity equation computations.	A plot of $G$ as a function of $\beta_{RC}$ effectively illustrates, in a single graph, the departure from equilibrium of a TBL. Such plots have been used for instance recently by \cite{knopp2021experimental}, \cite{fritsch2022fluctuating} and \cite{volino2023comparison}. We also use here $G$ as a function of $\beta_{RC}$ instead of $H$ as a function of $\beta_{ZS}$ for three reasons: $G$ is less sensitive to Reynolds number effects than $H$, the pressure gradient parameter that is compatible with $G$ from the similarity analysis is $\beta_{RC}$ (not $\beta_{ZS}$), plots of $G$ as a function of $\beta_{RC}$ are the ones commonly used.

	Figure \ref{ggg_beta} presents such a plot with the current flow case, together with three types of equilibrium or near-equilibrium data to appreciate its departure from equilibrium. The latter are the equilibrium TBL data of Mellor and Gibson (1966) from their similarity analysis, an empirical correlation from Cousteix (1989) based on near-equilibrium TBL data from the 1968 Stanford conference \citep{kline1969computation} that allows a comparison at high values of $\beta_{RC}$, and six cases of near-equilibrium APG TBLs from the literature \citep{kitsios2016direct,kitsios2017direct,bobke2017history,vila2020experimental,bradshaw1967turbulence,skaare1994turbulent,east1980investigation}. The Mellor and Gibson equilibrium data is probably reliable; however, a turbulence closure model had to be employed for its computation. The near-equilibrium flow case at $\beta_{RC}=34.2$ of \cite{kitsios2017direct} seems to deviate from the trend given by both the case of East and Sawyer (1980) at $\beta_{RC}=61.6$ and Cousteix’s empirical correlation, but it is impossible to know which dataset might be more biased. The departure from equilibrium of the present flow case can still be appreciated despite these discrepancies.
	
	Indeed, the response lag of the mean flow (mean velocity defect) can be seen throughout the streamwise evolution of the present case. At the start, the defect shape factor $G$ is very slightly lower than the equilibrium case for a given value of  $\beta_{RC}$, indicating that the momentum-losing effect of the pressure force is not immediate. At $\beta_{RC} \approx 6$, the pressure force starts decreasing in importance in the outer region which should lead to momentum gain. However, the momentum defect continues to increase substantially due to the influence of the upstream momentum-losing effect in the outer region and the increased impact of the pressure force in the inner region. The momentum defect even becomes superior to the equilibrium one at identical $\beta_{RC}$. Figure \ref{ggg_beta} illustrates that the Rotta-Clauser parameter $\beta_{RC}$ effectively reflects the momentum-losing effect of the pressure gradient across the entire boundary layer. The defect eventually starts decreasing, but it always remains higher than the equilibrium case. In the FPG zone at the end ($\beta_{RC}<0$), the defect remains significantly higher than that in equilibrium FPG TBLs.

	\begin{figure}
		\centering
		\includegraphics[scale=0.8]{ 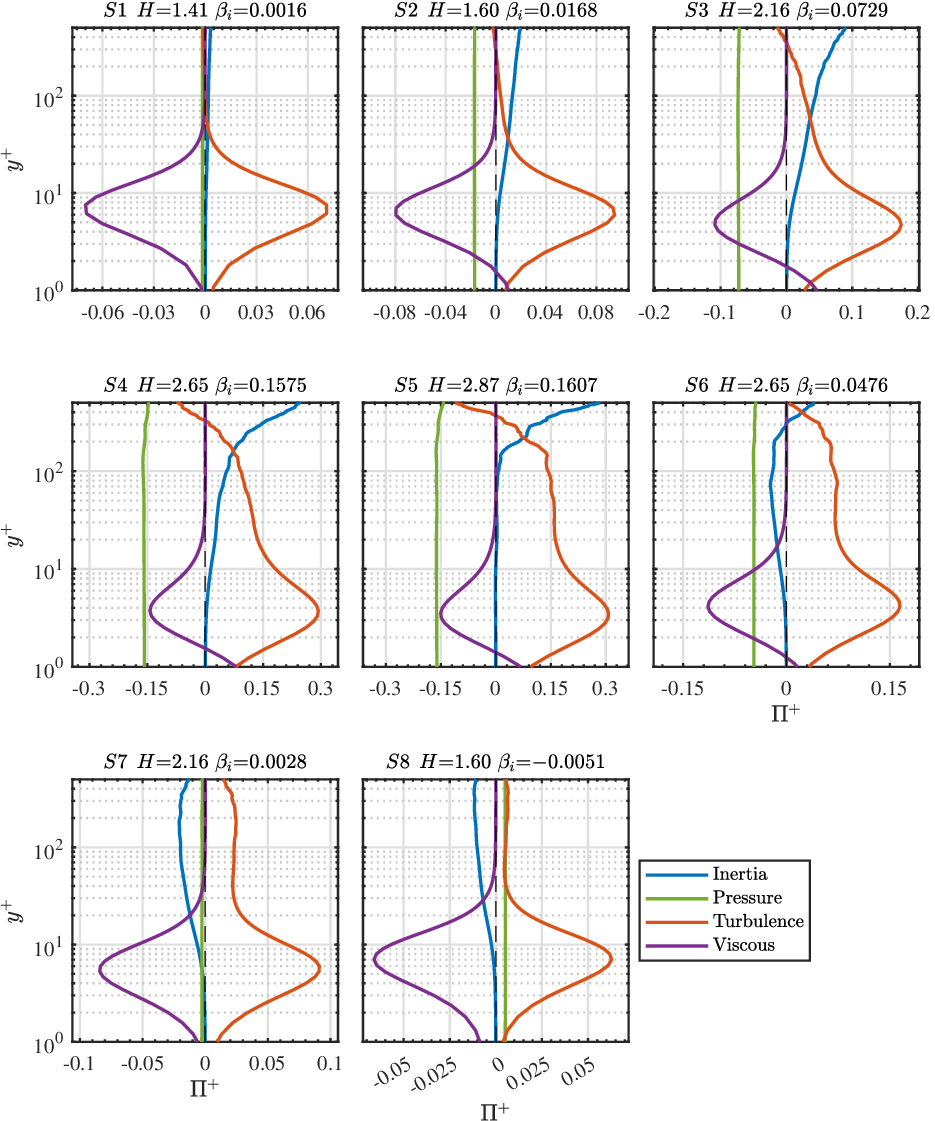}
		\caption{The mean momentum budget profiles of the eight streamwise positions as functions of $y^+$. The budget terms are normalized with friction-viscous scales. The $x$-axis ranges are chosen so that $F_{ti,max}$ is approximately at the same location in all sub-figures.}
		\label{mean_mom_inner}
	\end{figure}

	\subsubsection{The inner layer}

	As described in the introduction, $\beta_i$ represents the ratio of pressure force to turbulent force in the inner layer. Figure \ref{fig:chars}$(c)$ displays both $\beta_i$ and the ratio of pressure force to the maximum turbulent force in the inner layer, where $F_{p,w}$ is the pressure force at the wall. Although the distributions of force balance and $\beta_i$ do not follow each other as close as $\beta_{ZS}$ and the force balance in the outer layer, the trend is the same. $\beta_i$ shows the changes in force balance well and its behaviour is consistent with the behavior of $C_f$ as given in figure \ref{fig:chars}($d$). Figure \ref{mean_mom_inner} displays the force balance at the eight streamwise positions for the inner layer as a function of $y^+$. The $\beta_i$ distribution indicates that the pressure force's relative importance (as a momentum-losing force) in the inner layer increases until approximately $21\delta_{av}$ (between S4 and S5). The force balance from S1 to S4 also demonstrates this as the pressure force's relative importance in the inner layer with respect to the other forces increases and later decreases from S5 to S7, and becomes a positive force (FPG) at S8.
	
\begin{figure}
		\centering
		
		\includegraphics[scale=0.5]{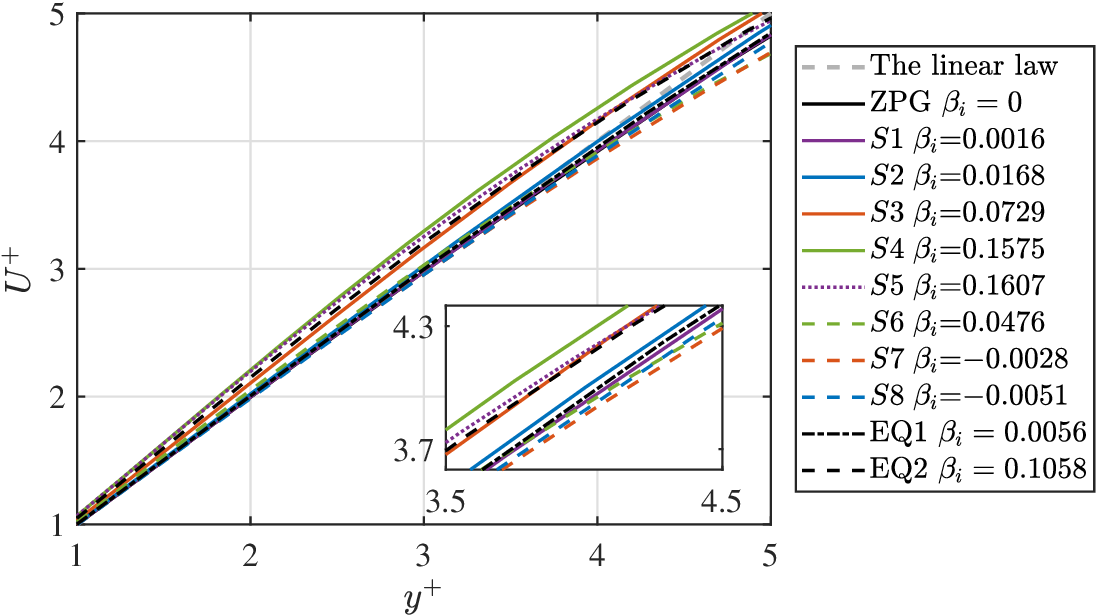}
		
		\caption{The friction-viscous scaled mean velocity profile of the eight streamwise positions, the near-equilibrium cases, and the ZPG TBL case in the viscous sublayer as a function of $y^+$. The black line (ZPG TBL) is hidden behind the purple line (S1) as they coincide within the viscous sublayer.}
		\label{upyp_zoom_vis}
	\end{figure}

	This change in force balance affects the mean velocity significantly in the inner layer. Figure \ref{upyp_zoom_vis} displays the mean velocity as a function of $y^+$ for the eight streamwise positions along with the ZPG TBL case in the viscous sublayer. Note that the ZPG TBL curve, in black, is hidden behind the purple curve of  S1 as both curves coincide in the viscous sublayer. As the pressure force's relative importance increases ($\beta_i$ increases), $U^+$ increases and deviates from the linear law. This increase is expected as it reflects the change from a linear law to a quadratic law at the wall when the pressure force increases in importance with respect to the wall shear stress force \citep{patel1973unified,skote2002direct}. This behaviour of $U^+$ in the viscous sublayer of APG TBLs was already reported before by \cite{gungor2016scaling}. The deviation from the ZPG TBL curve reaches its maximum near the position where $\beta_i$ is the highest. After $\beta_i$ reaches its maximum, $U^+$ constantly decreases, which means the pressure force's relative importance begins decreasing. Figure \ref{upyp_zoom_vis} shows that the mean velocity reacts almost instantly to the relaxation of the pressure force in the viscous sublayer. This immediate response of the mean flow to the decrease of the pressure force's relative importance suggests that the effect of flow history is small in the viscous sublayer. This is consistent with the fact that inertia forces, which are linked to the flow history's effect, are small in the viscous sublayer.

	There is still a deviation from the ZPG TBL profile in S6 to S8 which are located in the region where $d\beta_i/dx$ is negative, albeit the deviation is mild. The decrease of pressure force's importance from S6 to S8 leads to a decrease in $U^+$ in the viscous sublayer. In the last position, the pressure force is positive (FPG) which makes S8 different from all other positions. It is important to highlight that the flow would have not become a ZPG TBL if we had kept maintaining $d\beta_{ZS}/dx<0$ in a longer domain. The boundary layer would have eventually relaminarized.

\begin{figure}
		\centering
		\begin{tikzpicture}   
			\node(a){ \includegraphics[scale=0.4]{ 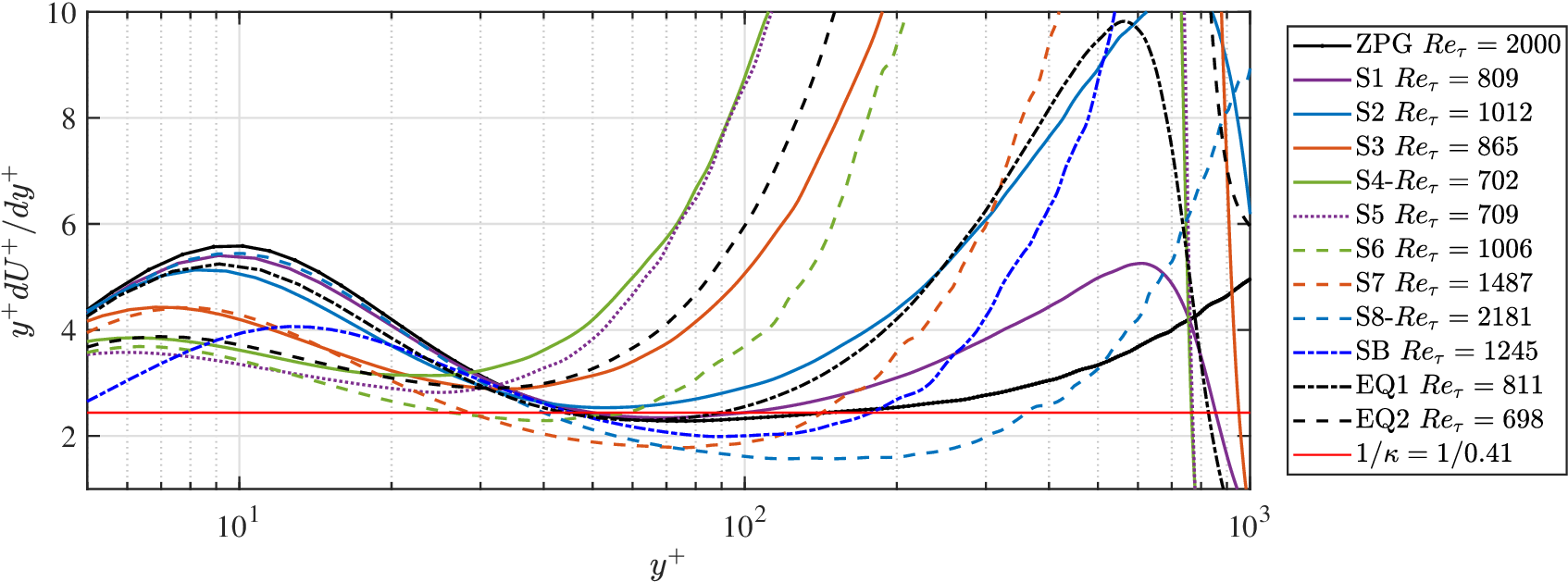}};
			\node at  (-5.75, 1.75) [overlay, remember picture] {($a$)};
		\end{tikzpicture} 
		\begin{tikzpicture}   
			\node(a){ \includegraphics[scale=0.4]{ 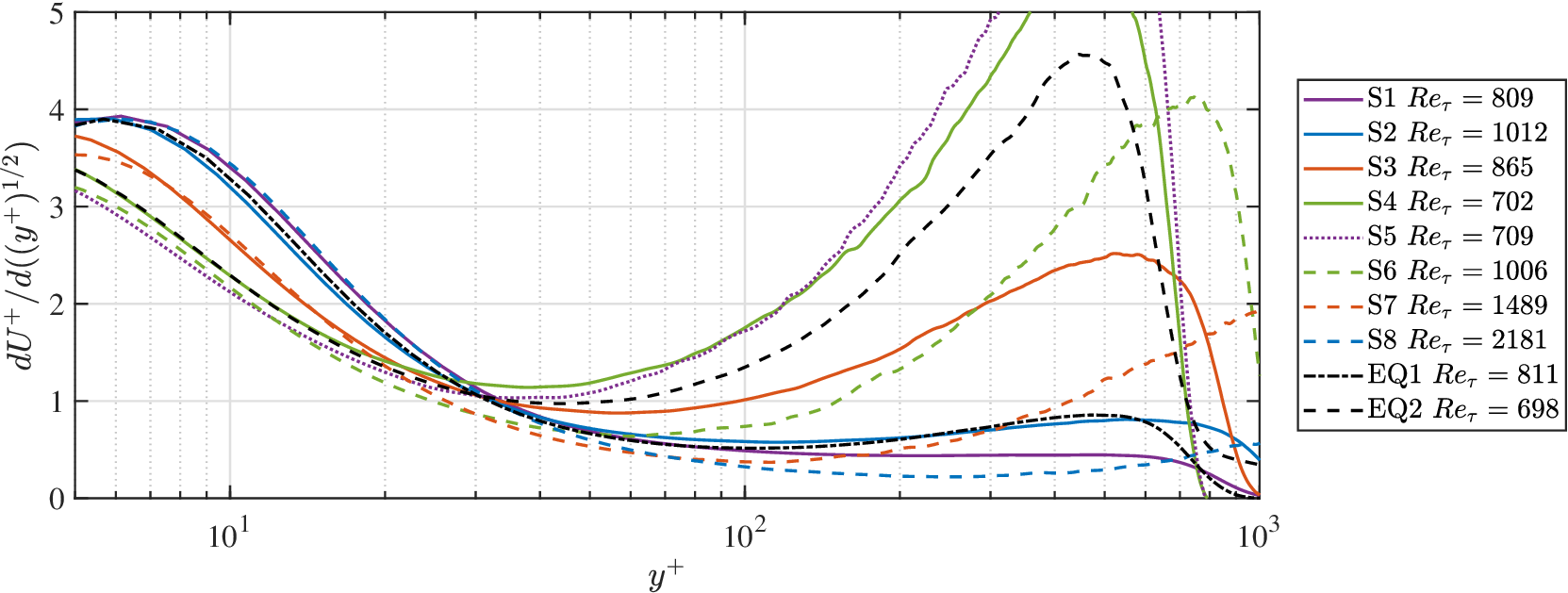}};
			\node at  (-5.75, 1.75) [overlay, remember picture] {($b$)};	\end{tikzpicture}                               
		\caption{ The diagnostic plots for the logarithmic law ($a$) and the half-power law ($b$) as a function of $y^+$. Straight red line in the top plot indicates the traditional Karman constant value $1/\kappa=1/0.41$.}
		\label{diagnostic}
	\end{figure}

\begin{figure}
		\centering
		\begin{tikzpicture}   
			\node(a){ \includegraphics[scale=0.4]{ 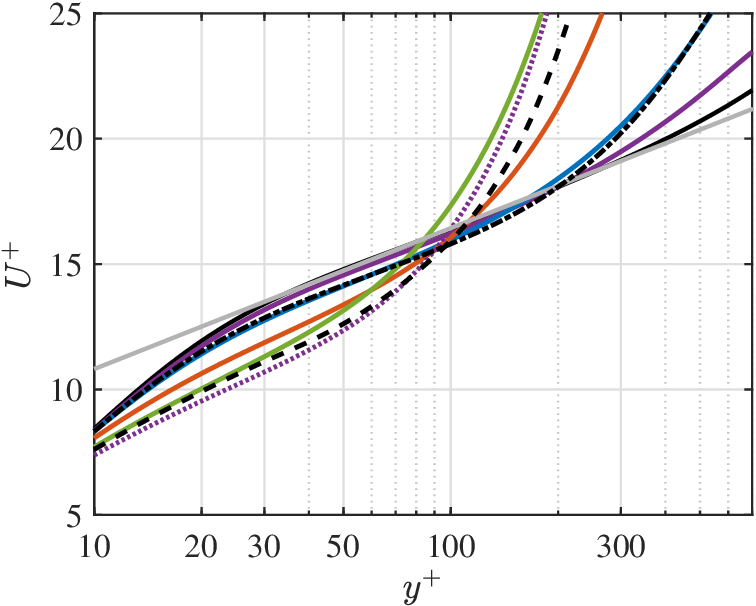}};
			\node at  (-1.5,1.6) [overlay, remember picture] {($a$)};
		\end{tikzpicture} 
		\begin{tikzpicture}   
			\node(a){ \includegraphics[scale=0.4]{ 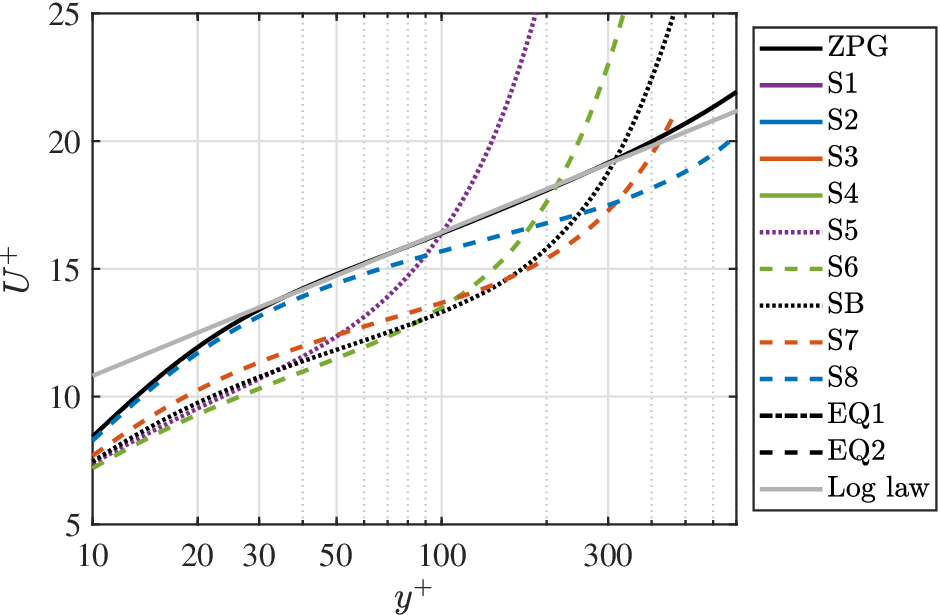}};
			\node at  (-2.2,1.6) [overlay, remember picture] {($b$)};
		\end{tikzpicture} 
		
		\begin{tikzpicture}   
			\hspace{-0.2cm}
			\node(a){ \includegraphics[scale=0.315]{ 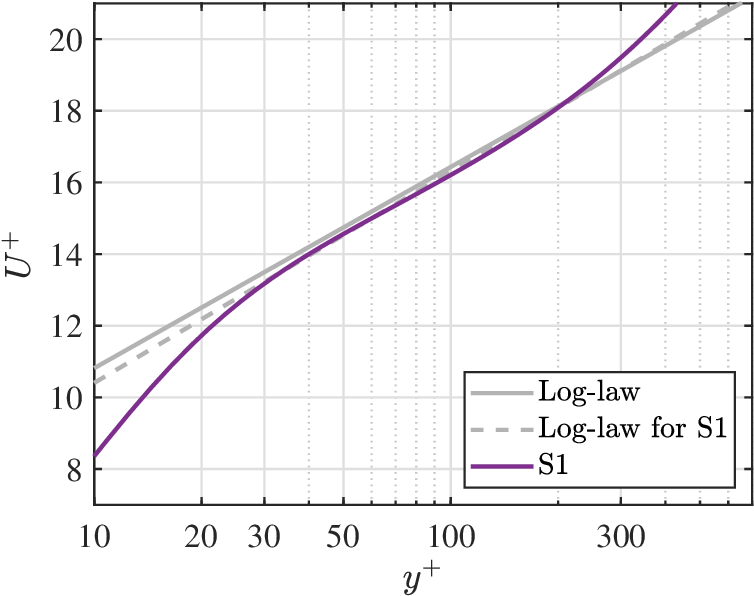}};
			\node at  (-1,1.2) [overlay, remember picture] {($c$)};
		\end{tikzpicture} \begin{tikzpicture}   
			\hspace{-0.2cm}                            
			\node(a){ \includegraphics[scale=0.315]{ 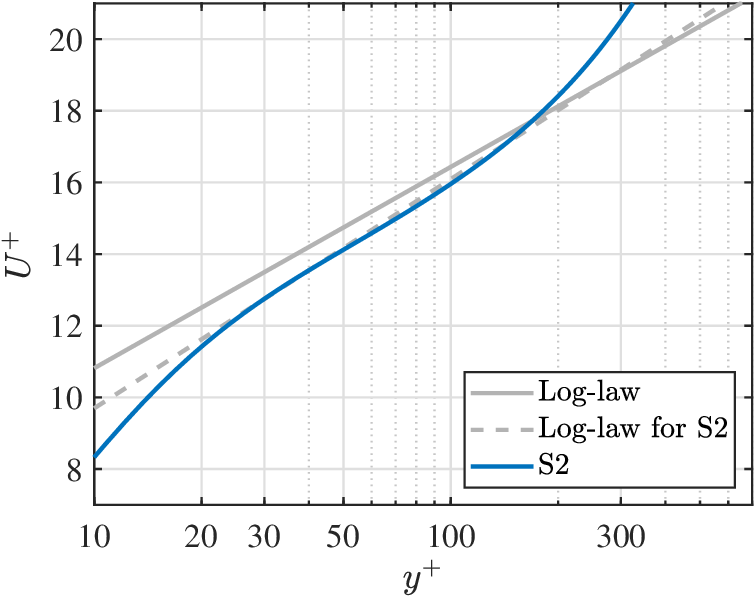}};
			\node at  (-1,1.2) [overlay, remember picture] {($d$)};
		\end{tikzpicture} \begin{tikzpicture}   
			\hspace{-0.2cm}
			\node(a){ \includegraphics[scale=0.315]{ 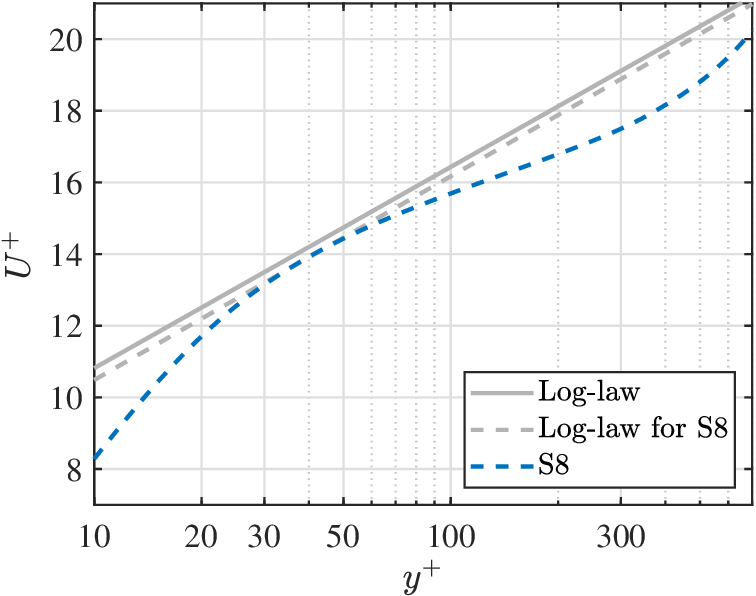}};
			\node at  (-1,1.2) [overlay, remember picture] {($e$)};
		\end{tikzpicture} 
		\caption{ Top: The mean velocity profile of various streamwise positions, the ZPG TBL case and and equilibrium cases in the overlap layer as a function of $y^+$. The streamwise positions from S1 to S5 are given on the left and the streamwise positions from S5 to S8 on the right for the sake of clarity. Bottom: The mean velocity profiles of three streamwise positions, S1 ($c$), S2 ($d$) and S8 ($e$) along with the logarithmic law using the traditional constants and values obtained with the correlations of \cite{nickels2004inner}, which were used in \cite{knopp2021experimental}.}
		\label{log_layer_figure}
	\end{figure}
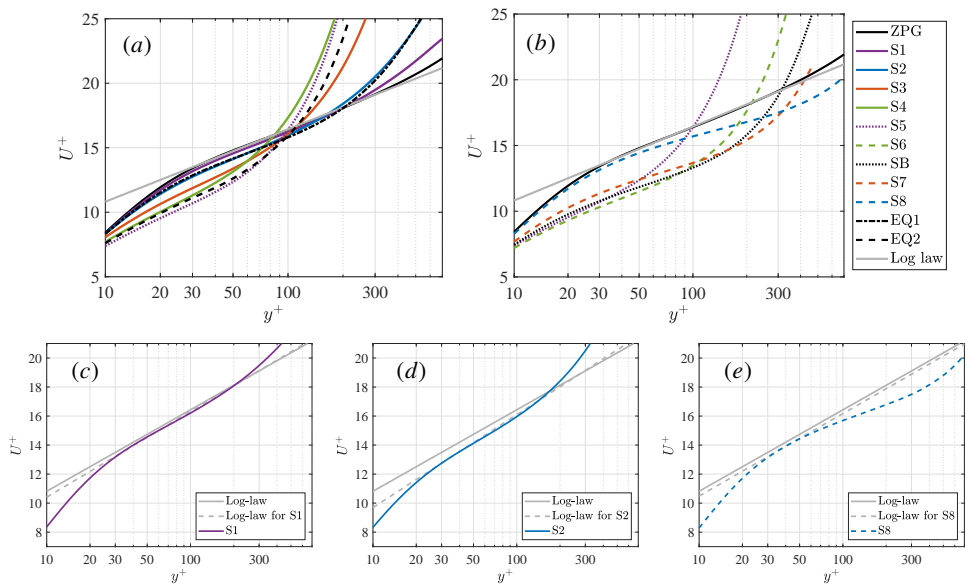

	Now, we discuss the overlap layer. It is important to state that there is no consensus regarding the nature of the overlap layer and the presence of the logarithmic law in APG TBLs in the literature, particularly when considering large-defect APG TBLs and those in disequilibrium. Furthermore, the overlap layer may be small or nonexistent due to the current flow being at a moderate Reynolds number. To investigate the potential presence of a logarithmic law, figure \ref{diagnostic}($a$) displays the diagnostic plot associated with such a law. The canonical logarithmic law does not hold, even in the small-defect case S1, as well as in the equilibrium cases EQ1 and EQ2. Nonetheless, the trends with respect to the logarithmic behaviour can be analyzed. Figures \ref{diagnostic} and \ref{log_layer_figure}$(c,d)$ show that the flow deviates from the logarithmic law with the traditional values of $\kappa$ and $B$, $0.41$ and $5$ respectively, even at S1 and S2 which are small-defect cases. Figures \ref{log_layer_figure}($a$) and \ref{log_layer_figure}$b$ illustrate that the consistently positive $\beta_i$ from S1 to S7 results in a drop of $U^+$ below the log law. Such an effect has been documented in numerous studies of non-equilibrium APG TBLs \citep{gungor2016scaling}. What is perhaps more surprising is its presence in the near-equilibrium cases depicted in figure \ref{log_layer_figure}($a$). This suggests that the phenomenon is, in part, a direct effect of the pressure gradient, rather than solely its disequilibrating nature (as indicated by an increasing $\beta_i$). In the case of three equilibrium APG TBLs with mild APGs, \cite{lee2017large} also observed a decreasing trend of $U^+$ below the log law with increasing $\beta_{RC}$, though these cases were at very low Reynolds numbers. Past experimental studies of near-equilibrium TBLs tended to confirm a log law behaviour \citep{clauser1954turbulent,east1980investigation,skaare1994turbulent,elsberry2000experimental}, even for cases with very large velocity defects, but since the wall shear stress was not measured directly, it is impossible to know if these flows follow the classical log law. Note that the large-defect near-equilibrium case EQ2 in Figure \ref{diagnostic}($a$) does not show a log law behaviour, which could be due to the moderate Reynolds number of the flow.

	\cite{knopp2021experimental} found similar deviations from the classical log law for an APG TBL with values of $H$ and $\beta_i$ close to those of S2 but with $Re_\theta$ an order of magnitude higher, above $20000$. But as they pointed out, the logarithmic law may exist with different values of $\kappa$ and $B$. Figures \ref{log_layer_figure}($c$) and \ref{log_layer_figure}($d$) show that the mean velocity at S1 and S2 approaches the logarithmic law obtained with the correlations of \cite{nickels2004inner} based on $\beta_i$, which were used in \cite{knopp2021experimental} for APG TBLs. The wall-normal extent of the approach towards a logarithmic law is thin in part because of the low Reynolds number. However, it is thinner in S2 than in S1 even though the Reynolds number is higher. This indicates that as the pressure force strength increases,  the logarithmic layer becomes thinner, eroding from above, as was found by Alving and Fernholz (1995), \cite{knopp2021experimental} and \cite{knopp2022empirical}.

	To illustrate the history effects on the mean flow in the overlap layer, we can consider another streamwise position (named SB) where $\beta_i$ has the same value $0.0167$ as S2 but with $H=2.40$ (between S6 and S7). The mean velocity profile of SB is shown in figure \ref{log_layer_figure}($b$), and its log-law diagnostic curve in figure \ref{diagnostic}($a$). The main differences between this position and S2 are the flow history and the sign of $d\beta_i/dx$ (positive in S2 and negative in SB). At SB, the flow does not follow the classical log law at all and it behaves completely differently from S2 due to upstream history effects. As another illustration that the non-equilibrium nature of the TBL transforms the overlap layer, $\beta_i$ is nearly zero at S7 but the mean velocity profile in the overlap region is very different from the ZPG TBL, where $\beta_i$ is nearly zero too. The behaviour of this non-equilibrium TBL clearly shows that it is not only the local balance of forces that matters, as reflected by $\beta_i$, but also the upstream history and the local type of disequilibrating effect of the pressure force ($d\beta_i/dx$).

	Because of these effects, the correlations of \cite{nickels2004inner} and \cite{knopp2021experimental} do not work for the other stations from S3 to S8. Figure \ref{log_layer_figure}($e$) shows the mean velocity profile and the logarithmic law for S8. The mean velocity profile seems to approach the modified logarithmic law between $y^+=30$ and $50$ but it cannot be considered as such because the profile is not logarithmic there as can be seen from figure \ref{diagnostic}$(a)$. The profile at S8 might actually approach a log law between $y^+=100$ and $200$. It should also be noted that the flow at S8 is an FPG TBL so if the decrease of $\beta_{ZS}$ (and $\beta_i$ and $\beta_{RC}$) would have been maintained in a longer domain, which means stronger FPG effects, the velocity profile would have eventually risen above the log law \citep{patel1968reversion,warnack1998effects,dixit2008pressure}. Such a rise above the log law was found in the transition from an APG TBL to an FPG TBL studied by \cite{tsuji1976turbulent}
	
	We also examine the presence of the half-power law which is expected to exist for large-defect TBLs \citep{stratford1959prediction,perry1966velocity,coleman2017direct,coleman2018numerical, knopp2021experimental}. Figure \ref{diagnostic}$(b)$ shows the diagnostic plot for the half-power law. It seems there is an approach towards the half-power law in large-defect cases S4, S5, and S6. The region where this approach occurs is located above the region where the mean flow approaches the logarithmic law as discussed previously. Therefore, the wall-normal location of the half-power region is consistent with the literature \citep{perry1966turbulent,knopp2021experimental}. As with the logarithmic layer, the half-power region is thin.

	\subsection{Turbulence}

	\begin{figure}
		\begin{tikzpicture}   
			\node(a){ \includegraphics[scale=0.45]{ 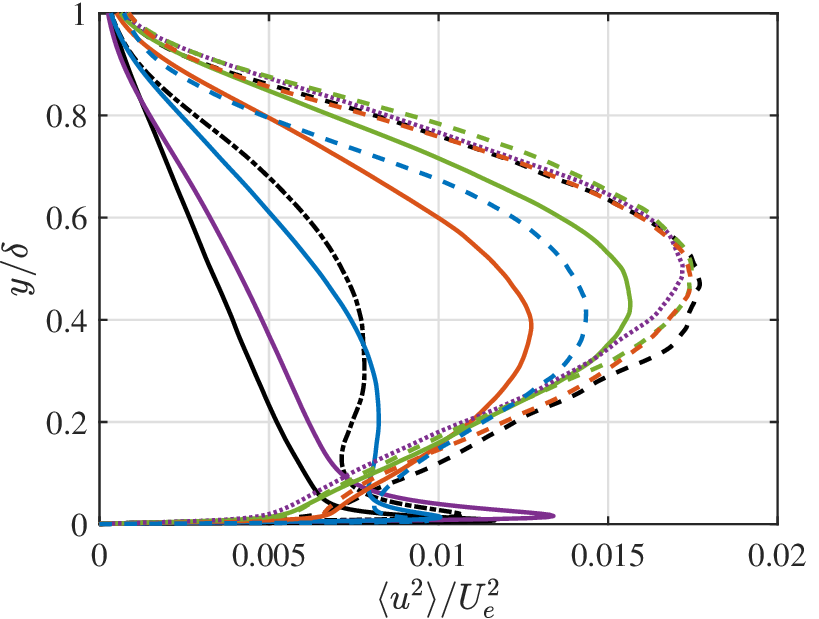}};
		\end{tikzpicture}  \hspace{0.5cm} 	\begin{tikzpicture}   
			\node(a){ \includegraphics[scale=0.45]{ 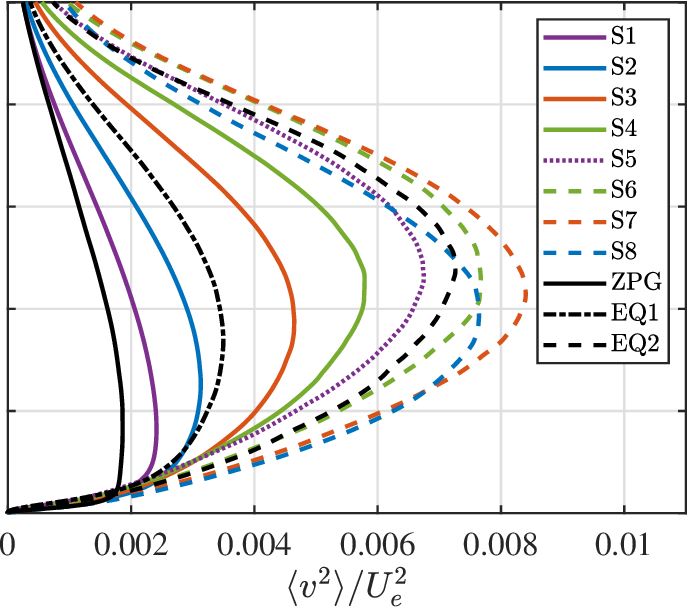}};
		\end{tikzpicture}  
		
		\begin{tikzpicture}   
			\centering                     
			\node(a){ \includegraphics[scale=0.45]{ 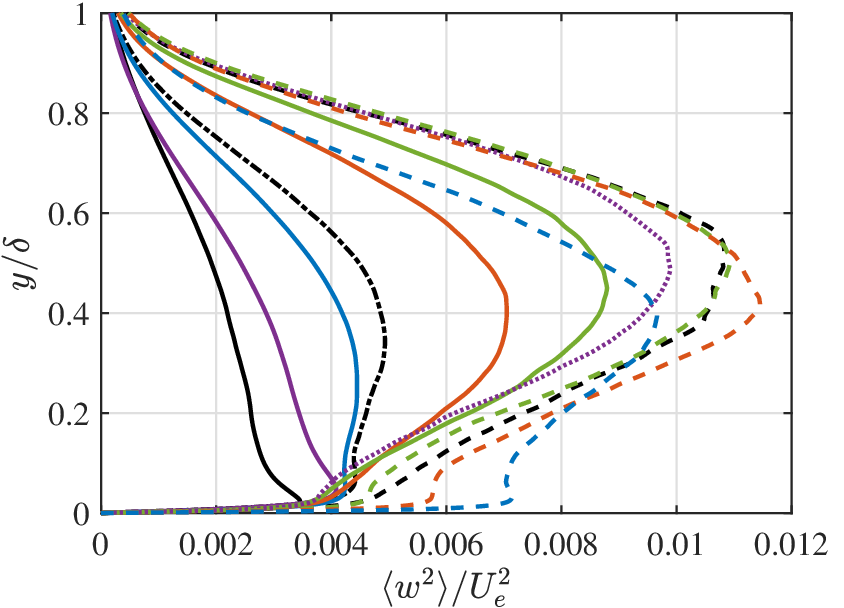}};
		\end{tikzpicture}   \hspace{-0.5cm}          \begin{tikzpicture}   
			\centering                     
			\node(a){ \includegraphics[scale=0.45]{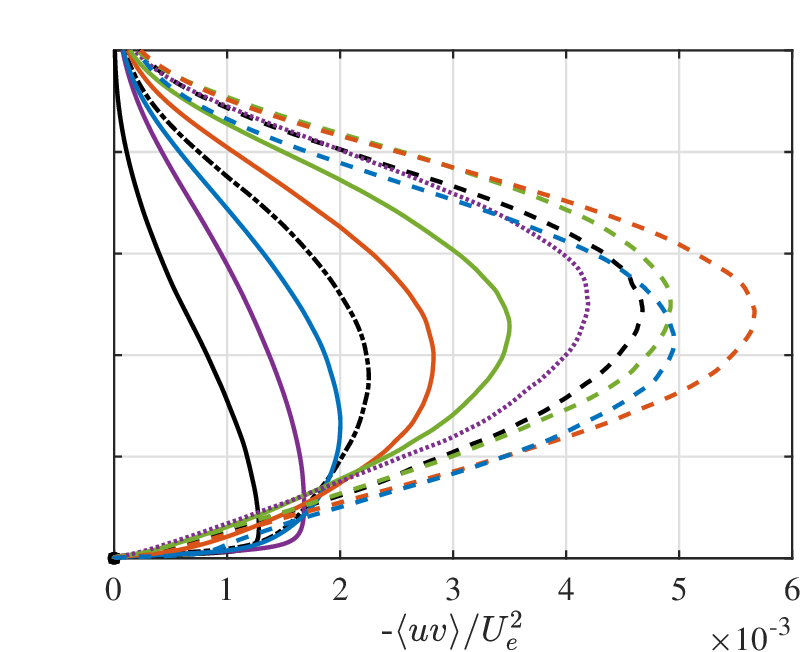}};
		\end{tikzpicture}  
		
		\caption{The Reynolds stress profiles normalized with the outer scales  as functions of $y/\delta$ for the eight streamwise positions, the ZPG TBL case and the near-equilibrium cases of EQ1 and EQ2.}
		\label{fig:u2ue}
	\end{figure}
	
	\subsubsection{Outer layer}

	The local and upstream pressure force effects on turbulence are now examined. Figure \ref{fig:u2ue} presents outer-scaled Reynolds stresses normalized with $U_e$. As it was reported numerous times for APG TBLs, outer-layer turbulence becomes dominant and inner-layer turbulence loses its importance as the velocity defect increases in the momentum-losing zone \citep{gungor2016scaling, maciel2018outer}. Moreover, the Reynolds stress levels in the outer layer increase with increasing velocity defect when they are normalized with $U_e$. However, the continuing rise of the Reynolds stresses while the flow is under turbulence-reducing conditions ($d \beta_{zs}/dx<0$), starting at S3, is indicative of the delayed response of turbulence to changes in the pressure force. This delayed response of turbulence is even more pronounced than that observed in the mean flow. As the velocity defect decreases in the outer layer, the mean shear also decreases (see figure \ref{fig:mean_vel}) and turbulence is expected to decay. But figure \ref{fig:u2ue} shows that outer turbulence keeps increasing all the way down to S7 due to the delay in response of turbulence to the changes in the pressure gradient and mean velocity. The delay in response is more pronounced for normal stresses $\langle v^2\rangle$ and $\langle w^2\rangle$, which continue to rise further downstream than  $\langle u^2\rangle$. This is reasonable since their energy depends on the redistribution from $\langle u^2\rangle$ through pressure-strain. The Reynolds shear stress follows the same trend as $\langle v^2\rangle$ and $\langle w^2\rangle$, with levels at S8 still higher than those of S5. It is important to remember that in an equilibrium FPG TBL or an FPG TBL monotonously evolving from a ZPG TBL, the Reynolds stresses in the outer region are lower than those of the ZPG TBL when scaled with $U_e$ \citep{harun2013pressure,volino2020non}. The Reynolds stress levels at S8 are therefore extremely high for an FPG TBL and this is due to the persistent effect of the upstream flow history.

	We now compare the near-equilibrium flow cases EQ1 and EQ2 with the present flow to isolate the effect of flow history. We first focus on the small defect cases: S2 and EQ1. Similar to our approach in analyzing the mean flow, we choose S2 because both cases exhibit a comparable velocity defect. The mean velocity profile, and consequently the mean shear profile, are similar in both cases. However, the local pressure force effect at S2 is stronger, as indicated by the values of $\beta_{ZS}$ and $\beta_{RC}$ in table \ref{table1}. Additionally, S2 is at a higher Reynolds number. Despite the similarities in the mean velocity profiles, the Reynolds stress profiles differ. There exists a disparity in the distribution of intensity among the Reynolds stress components. Specifically, the transverse Reynolds stresses ($\langle v^2\rangle$ and $\langle w^2\rangle$) exhibit higher intensity in the case of EQ1. Furthermore, the outer maxima of all Reynolds stress components are situated farther away from the wall in EQ1. In the current flow at S2, the pressure force has reached its local level through a ramp-up process. Consequently, the flow lacks the time needed to generate as much turbulence in the middle of the outer region compared to the near-equilibrium TBL. Moreover, there is insufficient time for the flow to redistribute as much energy to the transverse Reynolds stresses.

	\begin{figure}
		\centering 
		\begin{tikzpicture}   
			\node(a){ \includegraphics[scale=0.6]{ 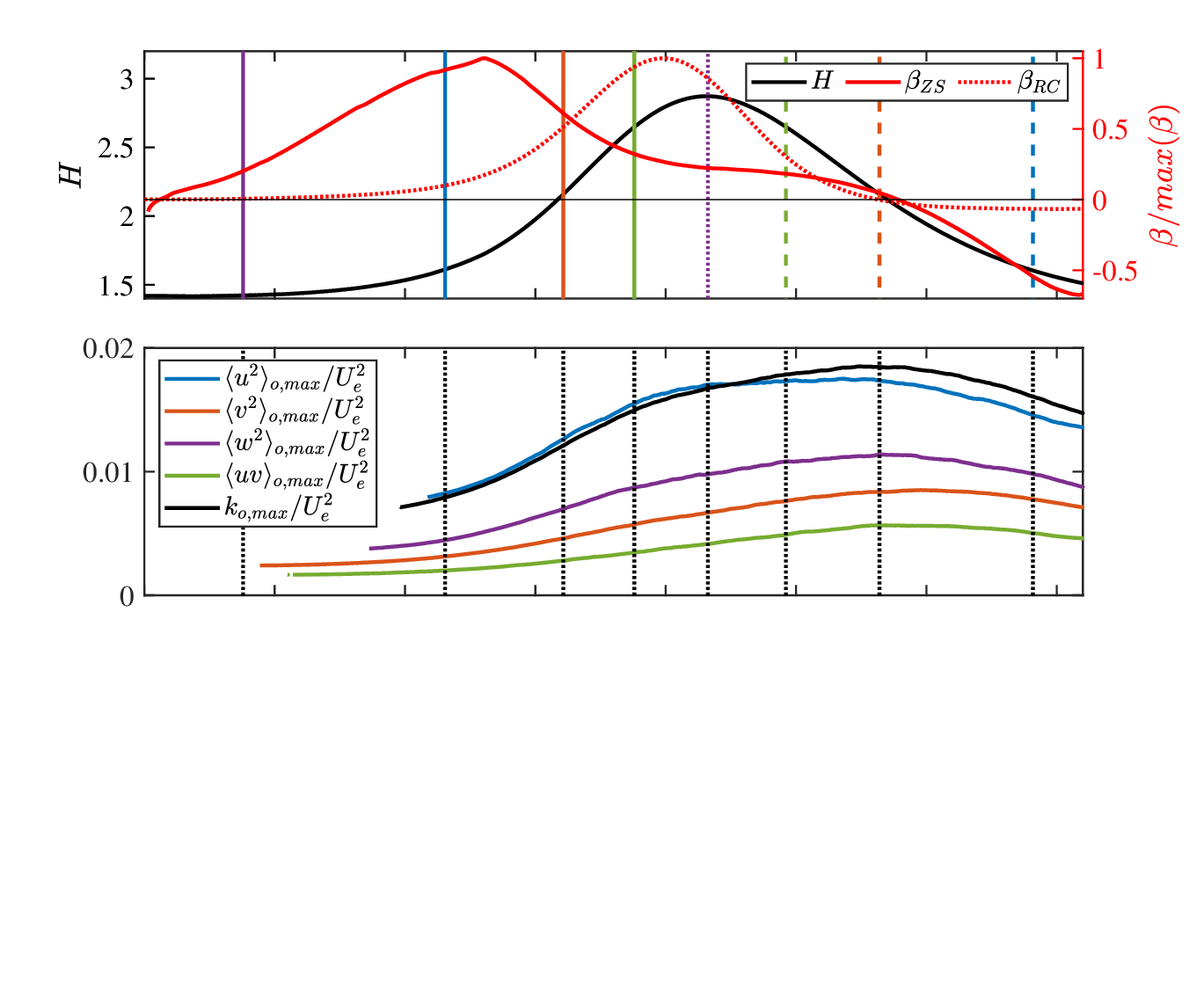}};
			\node at  (-4.7,2.8) [overlay, remember picture] {($a$)};		
			\node at  (-4.7,-0.7) [overlay, remember picture] {($b$)};
		\end{tikzpicture}  
		
		\vspace{-4.5cm}	
		\hspace{-0.5cm}	
		\begin{tikzpicture}   
			\node(a){ \includegraphics[scale=0.6]{ 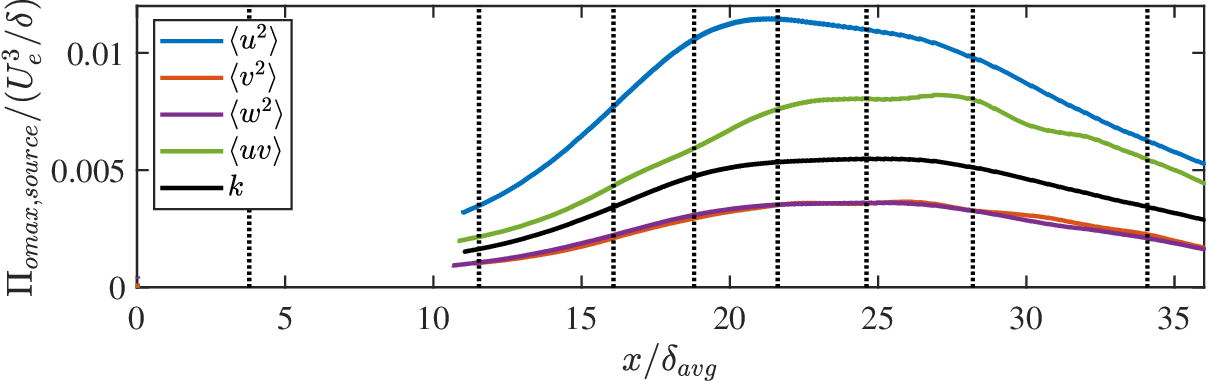}};
			\node at  (-4.4,-0.7) [overlay, remember picture] {($c$)};		
		\end{tikzpicture}

		\caption{The spatial development of $H$, $\beta_{ZS}/max(\beta_{ZS})$, and $\beta_{RC}/max(\beta_{RC})$  ($a$), the maxima of turbulent kinetic energy ($k$) and the Reynolds stresses ($b$) and source terms, which are production for $k$ and $\langle u^2\rangle$, and pressure-strain for $\langle v^2\rangle$ and $\langle w^2\rangle $ components ($c$) in the outer layer  as a function of $x/\delta_{av}$. Data only shown when an outer maximum is present. The levels are normalized with $U_e$ and $\delta$. }

		\label{fig:rsx}
	\end{figure}
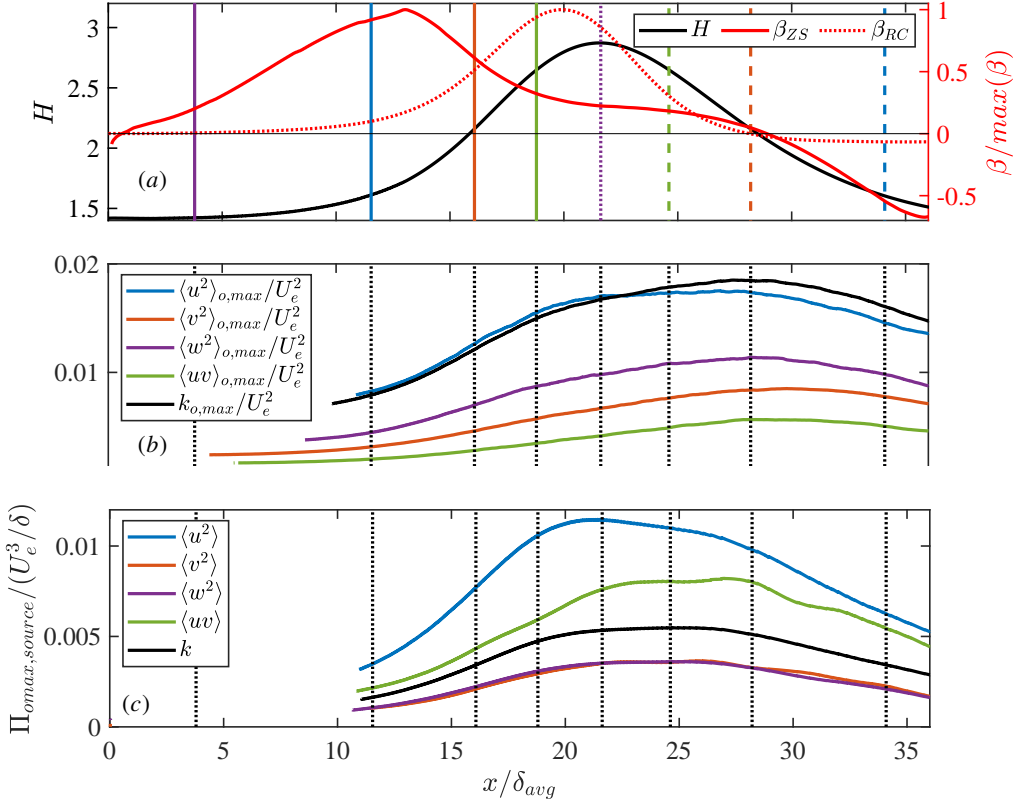
	
	Now, we compare the large defect flow cases, S4 and EQ2, which exhibit similar mean velocity profiles. In this case, the disparity in Reynolds stress levels is even more pronounced. The Reynolds stress levels at S4 are considerably lower than those in the near-equilibrium case.  However, unlike the small defect cases, the difference in the wall-normal positions of the outer maxima is less significant, although they still remain lower in the current non-equilibrium flow. The redistribution of energy to the transverse Reynolds stresses continues to be less important in the non-equilibrium flow.

	The delayed response of turbulence can be observed more comprehensively through the streamwise evolution of the outer maxima of the Reynolds stresses and turbulent kinetic energy, $k$, as depicted in figure \ref{fig:rsx}($b$). For ease of comparison with the pressure force effect and the mean flow response, figure \ref{fig:rsx}($a$) reproduces the streamwise distributions of $\beta_{ZS}$, $\beta_{RC}$, and $H$, also available in figure \ref{fig:chars}. The starting points of the curves in figure \ref{fig:rsx}($b$) correspond to the positions where the outer maxima of the Reynolds stresses begin to manifest. We have already seen that the mean flow responds with a delay, as evident from the $H$ distribution in figure \ref{fig:rsx}a, which is shifted downstream by approximately 10$\delta$ compared to the distribution of $\beta_{ZS}$. As mentioned earlier, the delay is even more significant for the Reynolds stresses in the outer region, and it varies among the different Reynolds stress components. The maxima of  $\langle u^2\rangle$ and $k$ peak at $x/\delta_{av} \approx 27$ and $28$, respectively, which is roughly 14$\delta$ downstream of the maximum of $\beta_{ZS}$. The delay is more pronounced for  $\langle v^2\rangle$ and even more so for  $\langle w^2\rangle$, again underscoring the delay in energy redistribution.
	
\begin{figure}
		\centering
		
		\includegraphics[scale=0.55]{  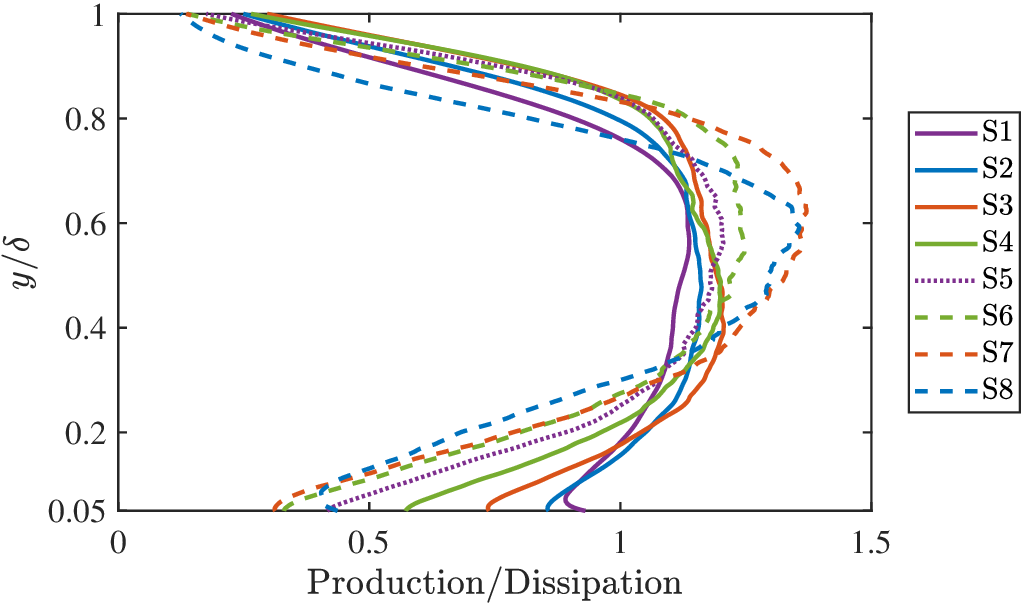}
		
		\caption{The ratio of production to dissipation of 	turbulent kinetic energy as a function of $y/\delta$. }
		\label{fig:ss_profiles}
	\end{figure}

	Outer turbulence therefore continues to increase for a considerable distance even after the pressure force begins to diminish. Furthermore, figure \ref{fig:rsx}($b$) illustrates that the subsequent decay of turbulence is slower than the initial upstream turbulence growth, even if the increase and decrease in pressure force are of the same amplitude (comparable $|d\beta_{ZS}/dx|$). This slow turbulence decay cannot be solely attributed to the prolonged survival of energetic turbulence structures convected from upstream, as this cannot explain the observed turbulence buildup where the pressure force is decreasing. A tentative explanation can be offered by analyzing the source and sink terms of $k$ and Reynolds stresses. Figure \ref{fig:rsx}($c$) illustrates the streamwise evolution of the maxima of the source terms of $k$ and the Reynolds stresses. In the case of $k$ and  $\langle u^2\rangle$, the source term is the corresponding production term while the main contribution comes from the pressure-strain (redistribution) term for $\langle v^2\rangle$ and $\langle w^2\rangle$. In addition, the  $\langle v^2\rangle$ production is considered for $\langle v^2\rangle$, albeit its contribution is small. It is observed in this figure that the production of $k$ and  $\langle u^2\rangle$ has peaked around S5. One might expect a subsequent decrease of  $k$; however, as noted earlier,  $k$ continues to increase up to $x/\delta_{av} \approx 28$ (figure \ref{fig:rsx}($b$)). This growth is attributed to an increase of the ratio of production to dissipation of $k$ in the middle of the boundary layer, from S5 to S7, as illustrated in figure \ref{fig:ss_profiles}. Indeed, figure \ref{fig:ss_profiles} demonstrates that the ratio of production to dissipation increases after S5 in the approximate wall-normal range $y/\delta=0.3$ to $0.8$. Given that this wall-normal range corresponds to the zone where production is maximal, the rise in the ratio of production to dissipation results in the continued increase of $k$ between S5 and S7 even if production has decreased in absolute terms.

	As a concluding exploration of outer region turbulence, we examine the budget of turbulent kinetic energy to gain a deeper understanding of turbulence behavior in the current non-equilibrium flow. The transport equation for the turbulent kinetic energy is given as follows.
	
	\begin{equation}\label{ters}
		\begin{split}
			0=&- 
			\bigg( U \frac{\partial k}{\partial x}+
			V \frac{\partial k}{\partial y} \bigg )   -  
			\bigg ( \frac{\partial U}{\partial x} \langle u^2 \rangle +  
			\frac{\partial U}{\partial y} \langle uv \rangle +
			\frac{\partial V}{\partial x} \langle uv \rangle +  
			\frac{\partial V}{\partial y} \langle v^2 \rangle \bigg ) \\ & 
			- \epsilon -	\frac{1}{2} \bigg ( \frac{\partial \langle u k \rangle}{\partial x} + \frac{\partial \langle v k \rangle}{\partial y} \bigg )
			+ 	 \upnu \bigg ( \frac{\partial^2 k}{\partial x^2}  + \frac{\partial^2 k}{\partial y^2}  \bigg ) 
			-	 \frac{1}{\rho} \bigg (\frac{\partial \langle pu \rangle}{\partial x}
			+ \frac{\partial \langle pv \rangle}{\partial y} \bigg )	\end{split}
	\end{equation}
	
\begin{figure}
		\centering
		\includegraphics[scale=0.6]{ 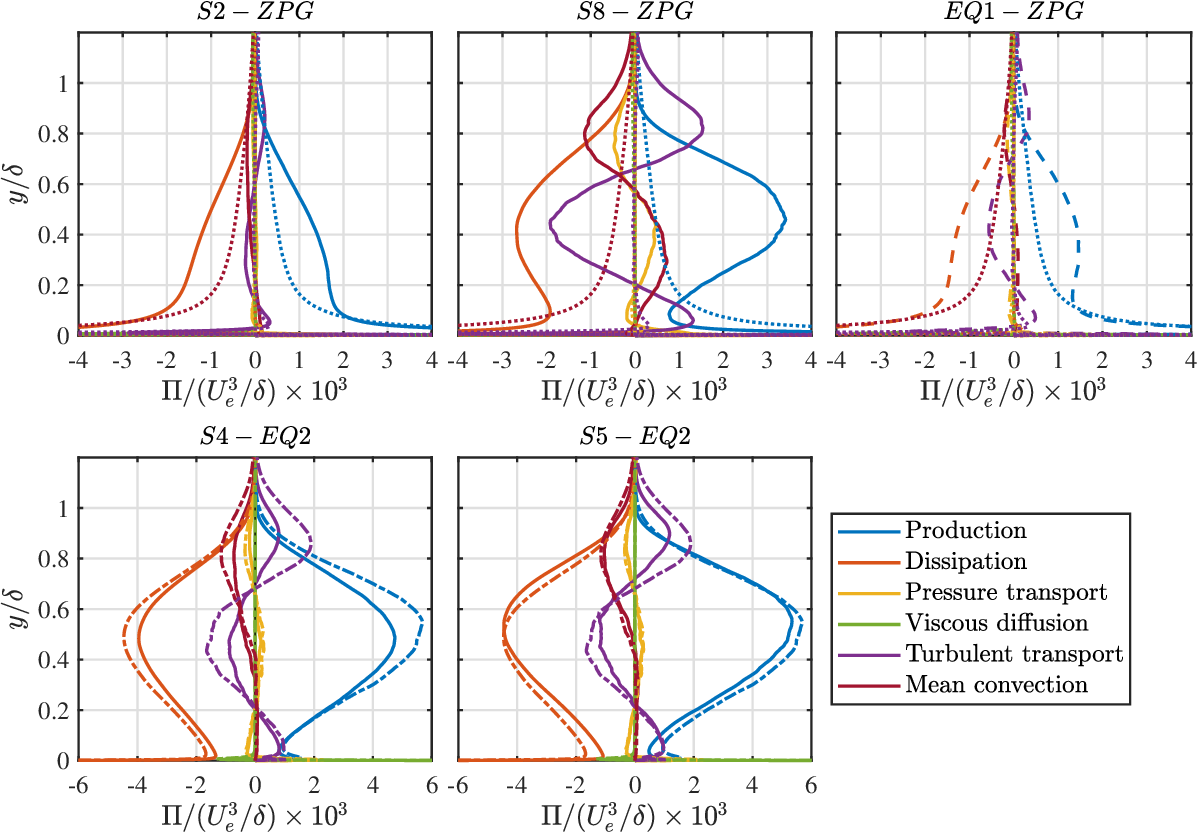}
		\caption{The turbulent kinetic energy budget for S2, S8, EQ1, ZPG, S4, S5 and EQ2 as a function of $y/\delta$. The terms are normalized with outer scales ($U_{e}$ and $\delta$). The straight, dashed, dashed-dotted, and dotted lines are for the present case, EQ1, EQ2, and ZPG, respectively. }
		\label{fig:outer_budget}
	\end{figure}

	\noindent The terms are in order, mean convection, production, dissipation, turbulent transport, viscous diffusion and pressure transport. The budgets are illustrated in figure \ref{fig:outer_budget}. For the sake of simplicity, for the current flow, we focus only on four representative stations that depict distinct flow situations: the small-defect case S2, the largest defect cases S4 and S5, and the small-defect case in the FPG region S8 which shares the same shape factor as S2. In each plot, a near-equilibrium case (ZPG, EQ1, or EQ2) is included as a reference for comparison.

	In the small-defect case S2, the turbulent kinetic energy budget, shown in figure \ref{fig:outer_budget}, resembles that of canonical wall flows as the comparison with the ZPG TBL indicates. However, there is an accumulation of production and dissipation in the outer region. This characteristic is typical of small-defect APG TBLs that either originated as ZPG TBLs (Gungor et al. 2022) or are in near-equilibrium state \citep{kitsios2016direct,kitsios2017direct,bobke2017history}. Nevertheless, when comparing the budget of S2 with that of the near-equilibrium case EQ1, which shares the same shape factor but has smaller pressure-gradient parameters $\beta_i$, $\beta_{ZS}$ and $\beta_{RC}$, the near-equilibrium case exhibits a local maximum of production around $y/\delta=0.35$, a feature absent in S2. The absence of a production maximum in S2 serves as another manifestation of the delay in the turbulence response of the current flow. Turbulent transport is also much stronger in the near-equilibrium case.
	
	The two lower plots compare the large-defect cases S4 and S5 with EQ2. In all these large-defect cases, there is a noticeable presence of production, dissipation, and turbulent transport in the middle of the boundary layer, consistent with the heightened turbulent kinetic energy observed in that region. The value of $\beta_{ZS}$ at S4 and EQ2 is almost the same, but the budget terms at S4 have not reached the levels observed in EQ2, once again indicating a delay in turbulence response. S5 is the case with the largest mean velocity defect. Even so, the budget terms are not more significant than those for EQ2. 
	
	The FPG TBL case S8 is compared with the ZPG TBL in the top-middle plot. Despite S8 being an FPG TBL, the outer maxima of the budget terms persist in the middle of the boundary layer due to the delayed response of turbulence, and their levels are much stronger than those in the ZPG TBL. An equilibrium FPG TBL with the same $\beta_{ZS}$ as S8 would exhibit a budget similar to the ZPG TBL, albeit with even lower levels in the outer region due to reduced mean shear. Although production and dissipation have decreased since S5, they remain elevated. Surprisingly, in contrast, turbulent transport at S8 is higher than at S5. Figure \ref{fig:u2ue} illustrates that the Reynolds stress profiles at S8 are wider than those at S5 due to these sustained high levels of turbulent transport.

	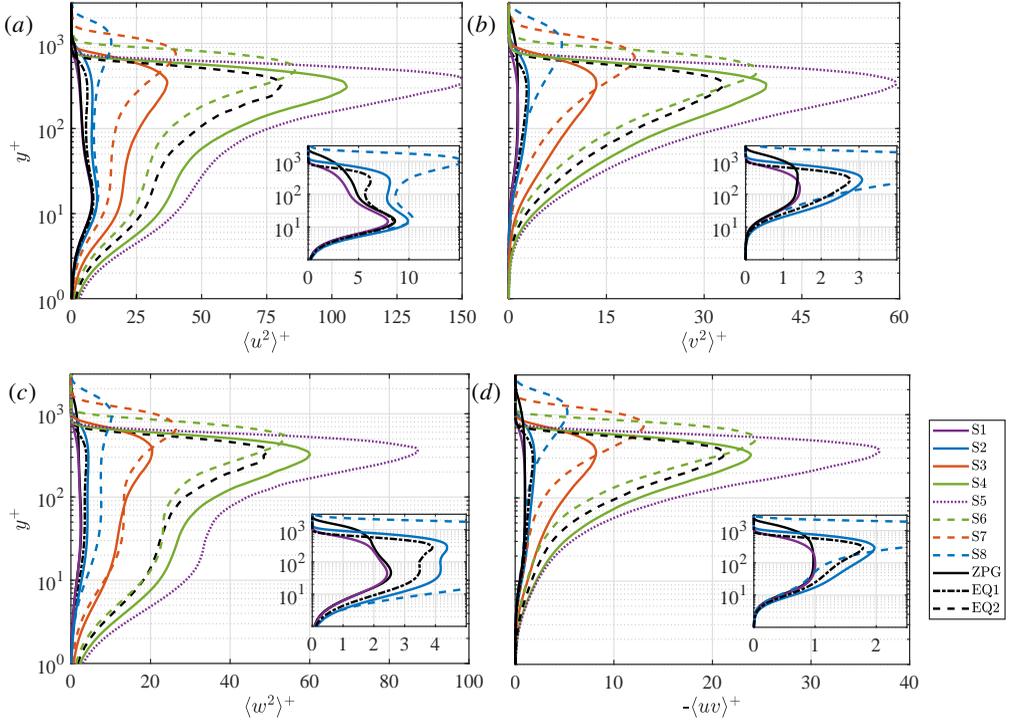
\begin{figure}
		\begin{tikzpicture}   
			\node(a){ \includegraphics[scale=0.45]{ 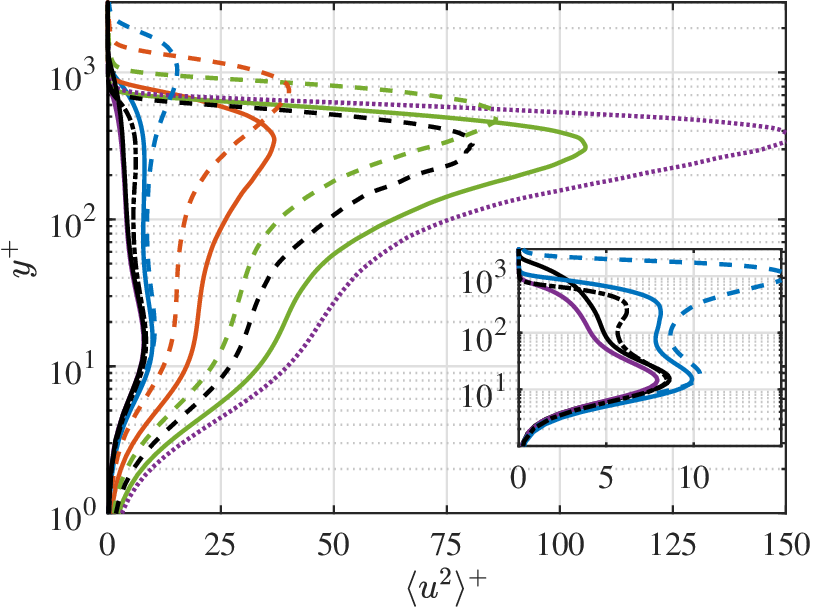}};
			\node at  (-3,2) [overlay, remember picture] {($a$)};			
		\end{tikzpicture}  
		\begin{tikzpicture}   
			\node(a){ \includegraphics[scale=0.45]{ 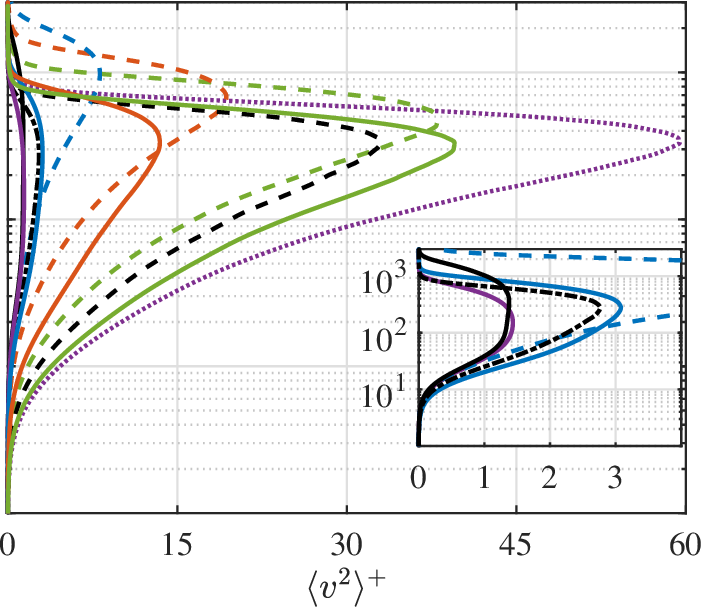}};
			\node at  (-3,2) [overlay, remember picture] {($b$)};			
		\end{tikzpicture}  
		
		\begin{tikzpicture}                 
			\node(a){ \includegraphics[scale=0.45]{ 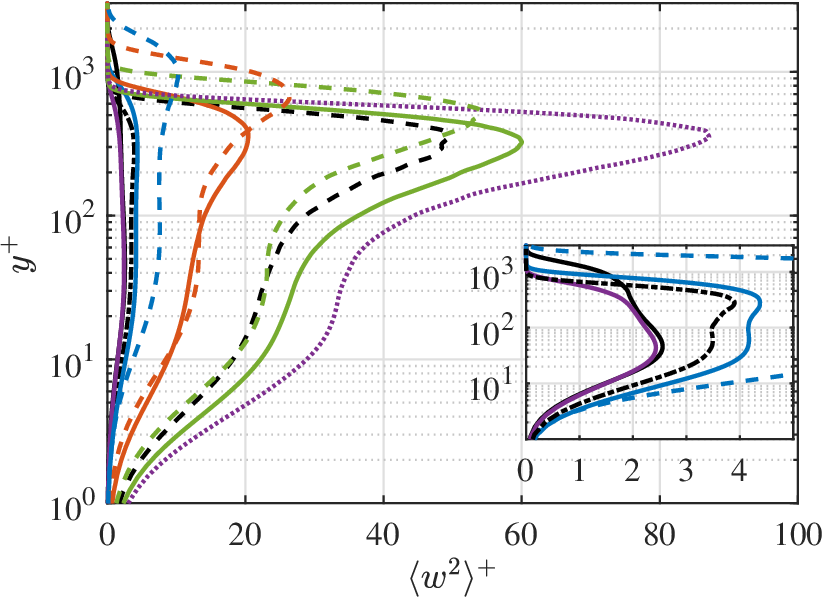}};
			\node at  (-3,2) [overlay, remember picture] {($c$)};			
		\end{tikzpicture}              
		\begin{tikzpicture}                 
			\node(a){ \includegraphics[scale=0.45]{ 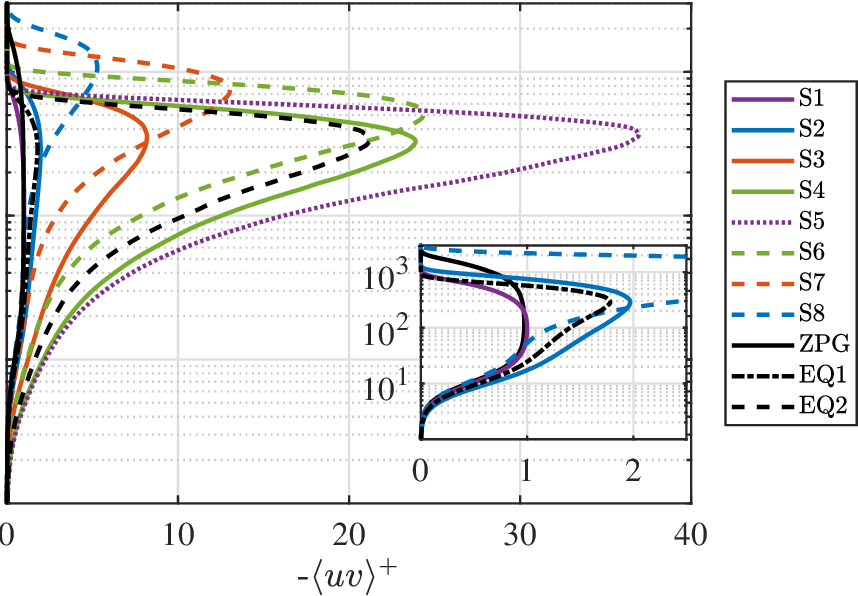}};
			Figure14d\node at  (-3.6,2) [overlay, remember picture] {($d$)};			
		\end{tikzpicture}  
		
		\caption{The Reynolds stress profiles normalized with friction-viscous scales as functions of $y^+$ for the eight streamwise positions, the ZPG TBL case and the near-equilibrium cases EQ1 and EQ2.}
		\label{fig:rs+}
	\end{figure}

	\subsubsection{Inner layer}

	Figure \ref{fig:rs+} illustrates Reynolds stress profiles as functions of $y^+$. When these profiles are normalized with friction-viscous scales, it is observed that the Reynolds stresses increase everywhere as $\beta_i$ and the velocity defect in the momentum-losing zone (until S5). This phenomenon has been well-documented in previous studies on APG TBLs \citep{gungor2016scaling,maciel2018outer}. The increase in Reynolds stress levels in such boundary layers occurs because the friction velocity is inadequate as a scaling parameter for Reynolds stresses when dealing with large velocity-defect TBLs \citep{maciel2018outer}. As the value of $\beta_i$ decreases, starting from S5 and continuing onwards, the amplitude of Reynolds stresses also decreases.
	
	In the near-wall region where inertia effects are negligible, below $y^+=10$, if turbulence were to respond instantly to the pressure force, the Reynolds stress profiles should be similar at identical $\beta_i$. However, stations S4 and S5, with close values of $\beta_i$, demonstrate that this is not the case. The Reynolds stress levels are higher at S5 than S4 in the near-wall region. Another indication of this discrepancy in response to the pressure force is the fact that the Reynolds stress profiles at S8 are higher than those in the ZPG TBL. Since $\beta_i$ is negative at S8 (FPG TBL), the Reynolds stress levels should be lower than those of the ZPG TBL in an equilibrium situation. These two observations suggest a delay in the response of turbulence. However, this delay may not be solely local; it could also be an indirect effect. As will be discussed later, part of the reason lies in the influence of large-scale turbulence on the near-wall region. The delayed response of the large-scale structures affects the near-wall region.
	
	\begin{figure}
		\centering
		
		\includegraphics[scale=0.45]{ 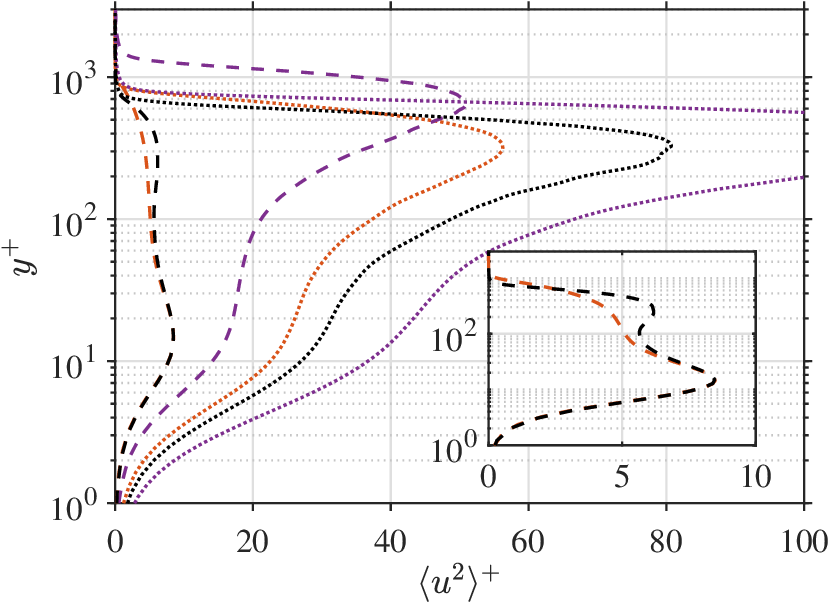}
		\includegraphics[scale=0.45]{ 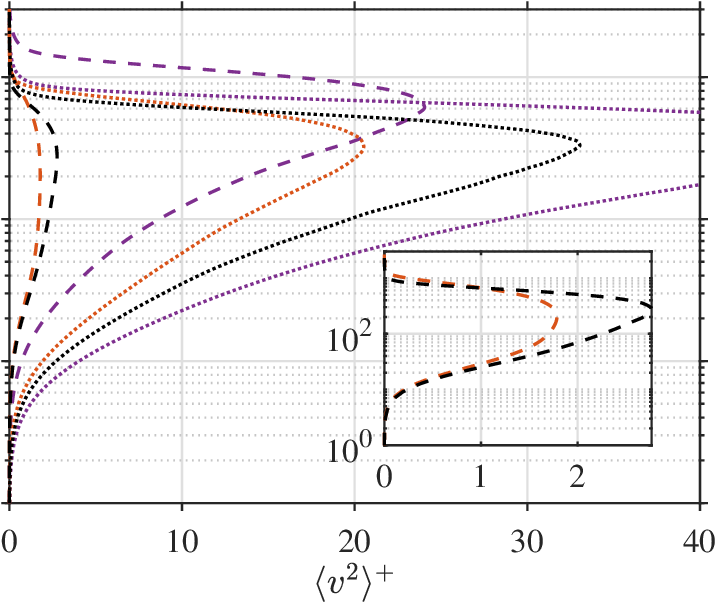}

		\includegraphics[scale=0.45]{ 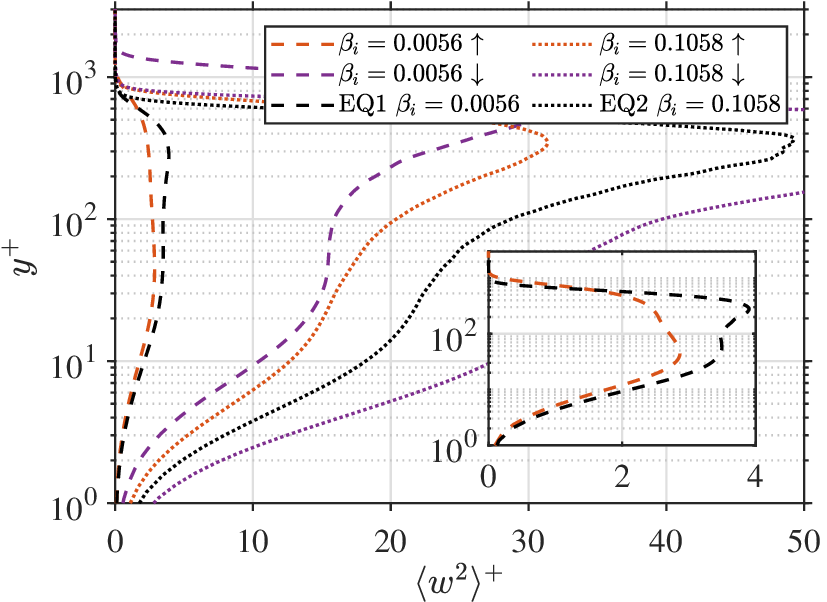}
		\includegraphics[scale=0.45]{ 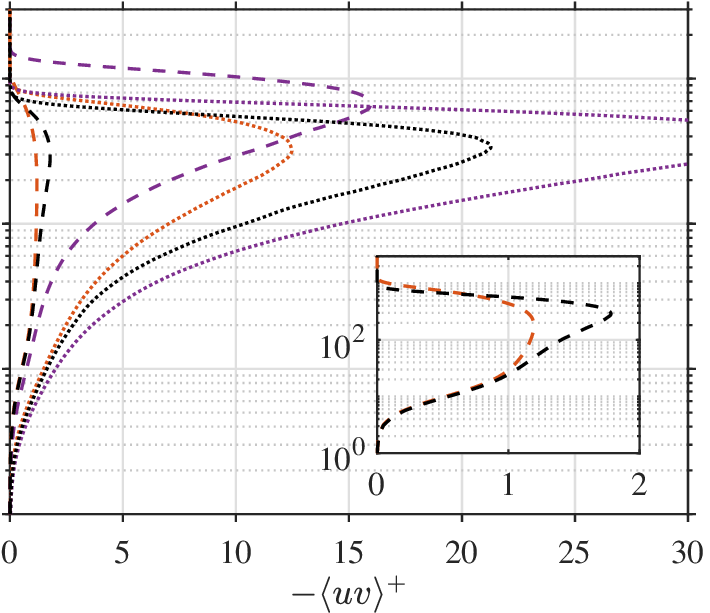}
		\caption{The Reynolds stress profiles normalized with friction-viscous scales as a function of $y^+$ for the streamwise positions of the present case along with EQ1 and EQ2 at matching $\beta_i$. The arrow indicates if $\beta_i$ is increasing or decreasing. 	}
		\label{fig:rs_betai}
		
	\end{figure}
	
	As for a comparison with the near-equilibrium cases, figure \ref{fig:rs_betai} illustrates the Reynolds stress profiles of the present flow, EQ1 and EQ2 at matching values of $\beta_i$. For the current flow, two profiles are shown at each $\beta_i$ value: one closer to the beginning of the domain, where the APG turbulence-promoting effect is increasing ($d\beta_i/dx>0$) and depicted in orange, and another further downstream where the APG effect is decreasing ($d\beta_i/dx<0$) and shown in purple. For the small-defect cases, a comparison between EQ1 and the first position in the current flow (dashed orange) reveals, in the zoomed-in plots of figure \ref{fig:rs_betai}, that the Reynolds stresses in the inner region match up to approximately $y^+=30$, 10 and 20 for $\langle u^2\rangle^+$, $\langle v^2\rangle^+$ and $\langle uv\rangle^+$, respectively. For $\langle w^2\rangle^+$, the profiles start diverging very near the wall, and no explanation has been found for such distinctive behavior. Overall, in the present flow, the Reynolds stresses in the inner region have not yet attained the levels observed in the near-equilibrium case EQ1. Nevertheless, these history effects remain relatively minor when compared to all other stations located further downstream. 
	
	Indeed, at the second small-defect station (dashed orange), much further downstream, the deviations from the near-equilibrium case are significant everywhere, even very near the wall. As for the large defect cases, the Reynolds stresses in the inner region at the first position in the current flow (dotted orange) have not reached the levels of the near equilibrium case EQ2. Meanwhile, at the second station (dotted purple), they have exceeded those levels. This underscores again the substantial delay in turbulence response to changes in the pressure force.

	Turning our attention to $\langle u^2\rangle^+$, figure \ref{fig:rs+}($a$) demonstrates that the near-wall peak, observed at $y^+=15$ in the ZPG TBL, becomes less sharp and eventually disappears as the local APG pressure force effect increases. This phenomenon, evident in both non-equilibrium and near-equilibrium flows, is attributed to the significant increase in $\langle u^2\rangle^+$ above the near-wall region. The vanishing of the near-wall peak, coupled with the rise of $\langle u^2\rangle^+$, has been previously documented (Nagano et al. 1998, Drodz et al. 2015, Gungor et al. 2016). This suggests that the small-scale near-wall turbulence activity might be obscured by the footprints of the energetic large-scale structures located above the near-wall region. To investigate this, we analyze the spectral content of $\langle u^2\rangle$ and $\langle uv\rangle$, along with wall-normal profiles of these Reynolds stresses decomposed into small and large scales. 
	
\begin{figure}
		\begin{tikzpicture}   
			\node(a){ \includegraphics[scale=0.8]{ 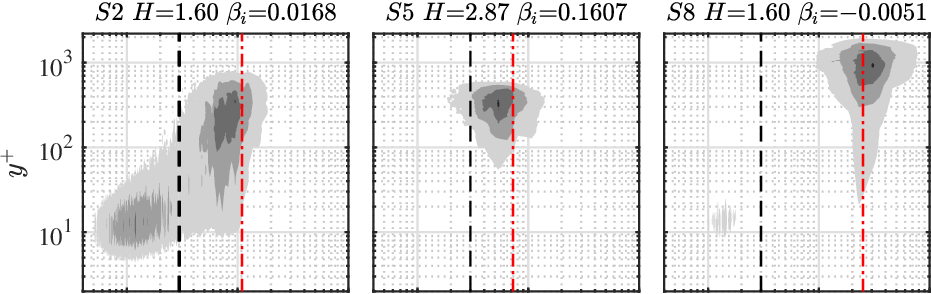}};
		\end{tikzpicture}  
		\begin{tikzpicture}   
			\node(a){ \includegraphics[scale=0.8]{ 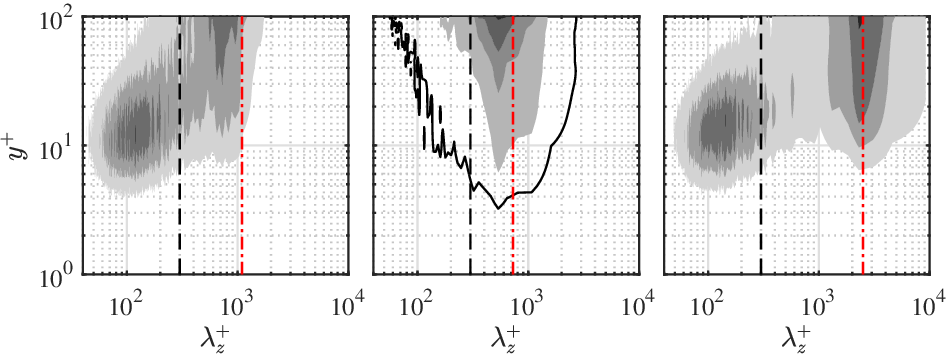}};
		\end{tikzpicture}

		\caption{The spectral distribution of $\langle u^2\rangle$ as a function of $\lambda_z^+$ and $y^+$ for S2, S5, and S8. The spectra are plotted for the whole wall-normal range (top) and for the region up to $y^+=100$ (bottom). The flooded-contour levels are 0.3, 0.5, 0.7, and 0.95 of the maxima of each spectra. The black-line contour is 0.1 of the maxima. The dashed black lines indicate the cutoff filter $\lambda_z^+=300$ and the red dash-dotted lines indicate $\lambda_z^+=\delta^+$.}
		\label{fig:spectra_u2}
	\end{figure}

\begin{figure}
		
		\begin{tikzpicture}   
			\node(a){ \includegraphics[scale=0.8]{ 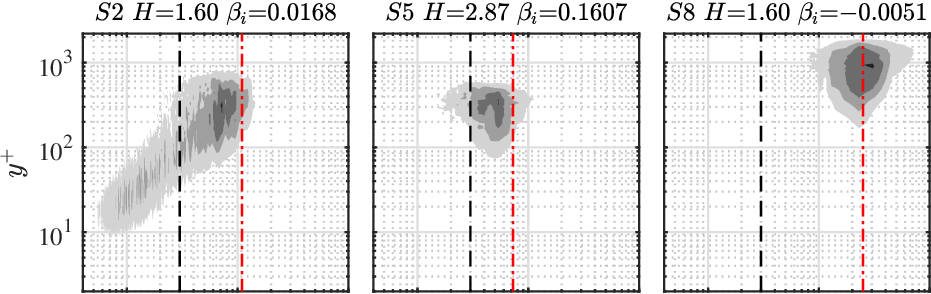}};
		\end{tikzpicture}  
		\begin{tikzpicture}   
			\node(a){ \includegraphics[scale=0.8]{ 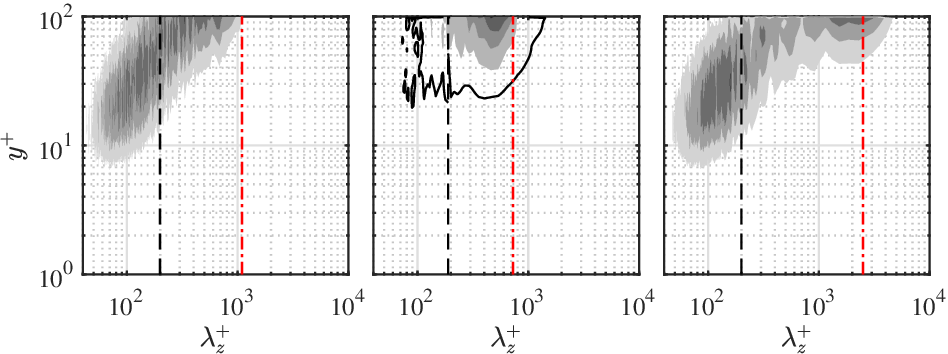}};
		\end{tikzpicture}  
		\caption{The spectral distribution of $\langle uv\rangle$ as a function of $\lambda_z^+$ and $y^+$ for S2, S5, and S8. The spectra are plotted for the whole wall-normal range (top) and for the region up to $y^+=100$ (bottom).  The flooded-contour levels are 0.3, 0.5, 0.7, and 0.95 of the maxima of each spectra. The black-line contour is 0.1 of the maxima. The dashed black lines indicate the cutoff filter $\lambda_z^+=300$ and the red dash-dotted lines indicate $\lambda_z^+=\delta^+$.}
		\label{fig:spectra_uv}
	\end{figure}

	Figures \ref{fig:spectra_u2} and \ref{fig:spectra_uv} depict the pre-multiplied $\langle u^2\rangle$ and $\langle uv\rangle$ one-dimensional spanwise spectra, respectively, as functions of $\lambda_z^+$ and $y^+$ for S2, S5, and S8. The black dashed lines in these figures delineate the wavenumber sharp filter cutoff that will later be employed to decompose small and large scales. The bottom row presents the same spectra as the top row but only up to $y^+=100$ and with chosen contour values to facilitate clearer visualization of the energy content in the inner region.

	The spectral distribution of $\langle u^2\rangle$ at S2 exhibits an inner peak, indicative of the near-wall streaks. These streaks are narrow structures with a $\lambda_z^+$ of approximately 120, a characteristic extensively discussed in the literature \citep{smith1983characteristics}. As the velocity defect increases from S2 to S5, the outer-layer turbulence becomes dominant, and the inner peak vanishes from the spectra. This is consistent with the literature for large-defect TBLs \citep{lee2017large,kitsios2017direct,gungor2022energy}. Despite the absence of the inner peak, it remains unclear whether the streaks have vanished or if they are obscured by the presence of more intense large-scale structures \citep{gungor2023coherent,gungor2024turbulent}.

	Following the reversal to momentum-gaining conditions induced by the pressure gradient effect ($d\beta_{ZS}/dx<0$ and $d\beta_i/dx<0$), the inner peak re-emerges at S8 with a similar $\lambda_z^+$, but it is weak compared to the outer peak. This finding is significant, demonstrating that inner-layer turbulence is becoming like that of canonical flows, as expected for an FPG TBL with a moderate $\beta_i$ value, despite the presence of history effects. It is important to emphasize once again that the flow is not reverting to a ZPG TBL state, as $d\beta_{ZS}/dx<0$ is maintained. If the pressure force variation were imposed in a longer domain, the flow would eventually laminarize. 
	
	In the inner region, $\langle u^2\rangle$ is also associated with wide structures at all stations. These large-scale structures have a $\lambda_z$ on the order of $\delta$, as in the outer region. At S5, these structures carry the majority of $\langle u^2\rangle$ in the inner layer. Furthermore, at S8, the relative contribution of the large scales in the inner region is higher than at S2 due to the presence of very large and energetic structures convected from upstream. This history effect also results in the structures being wider, with respect to $\delta$, at S8 than S2. \cite{gungor2024turbulent} showed that these large-scale structures primarily belong to the outer layer.

	\begin{figure}
		\begin{tikzpicture}   
			\node(a){ \includegraphics[scale=0.45]{ 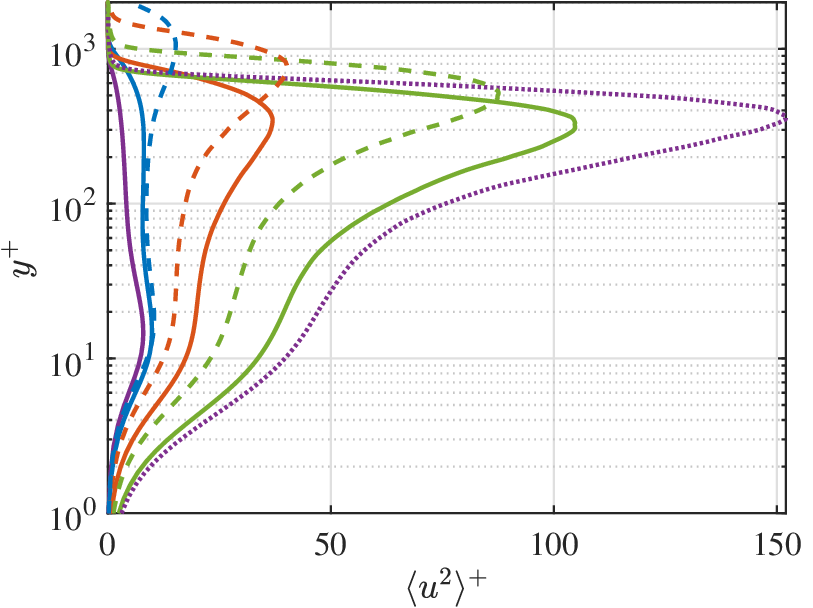}};
		\end{tikzpicture}  
		\begin{tikzpicture}   
			\node(a){ \includegraphics[scale=0.45]{ 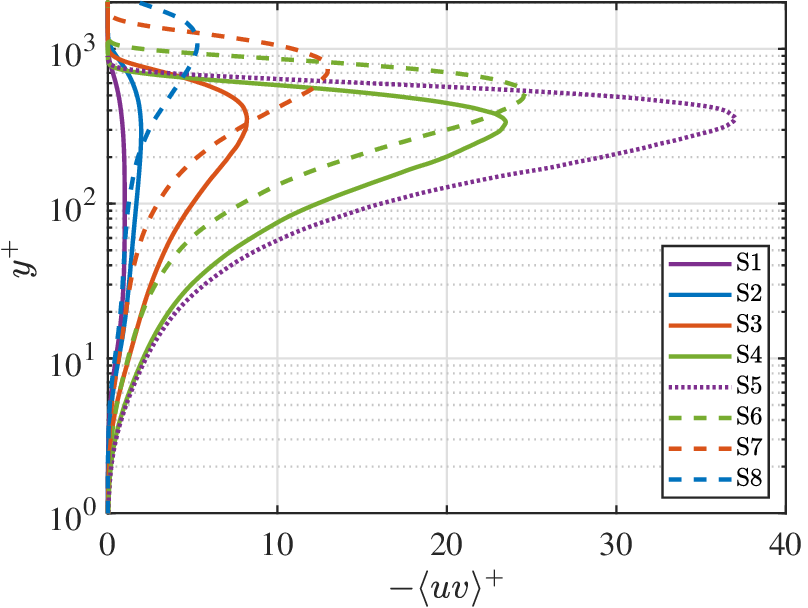}};
		\end{tikzpicture}  
		
		\begin{tikzpicture}   
			\node(a){ \includegraphics[scale=0.45]{ 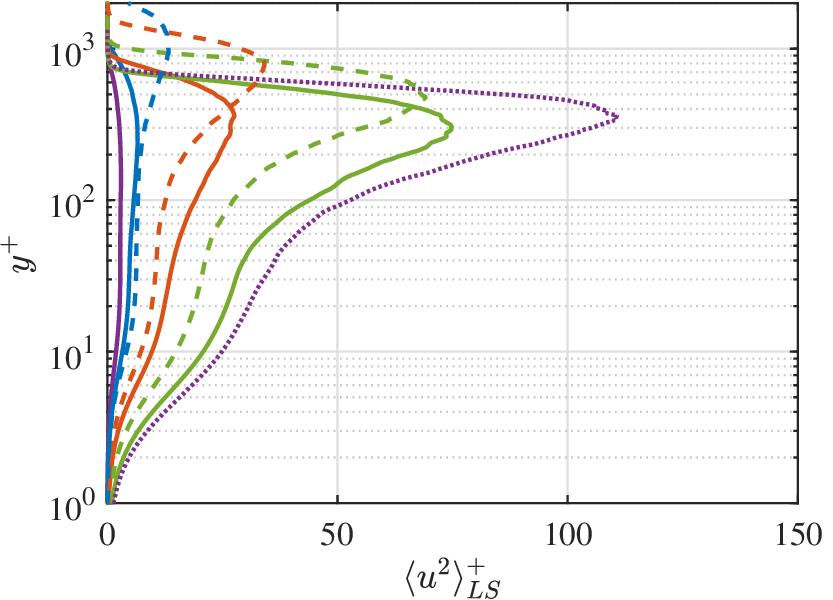}};
		\end{tikzpicture}  
		\begin{tikzpicture}   
			\node(a){ \includegraphics[scale=0.45]{ 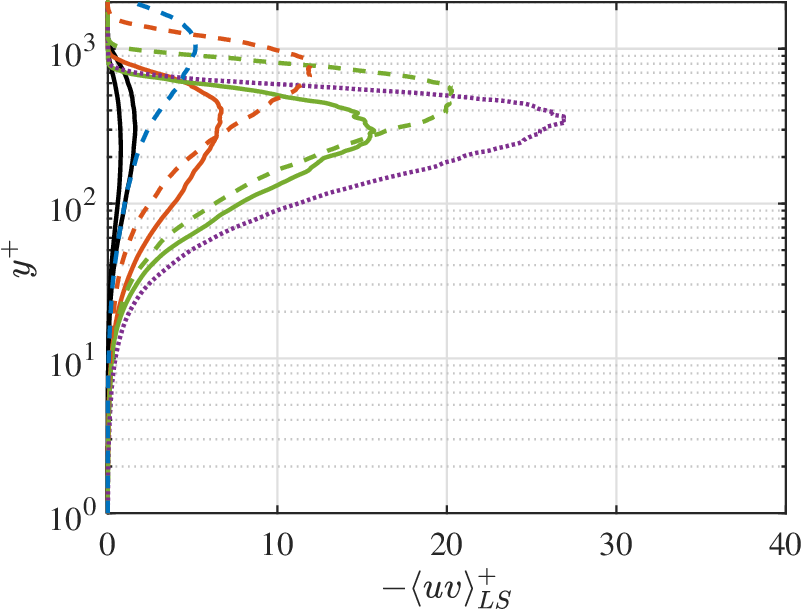}};
		\end{tikzpicture}

		\hspace{-0.05cm}		\begin{tikzpicture}   
			\node(a){ \includegraphics[scale=0.45]{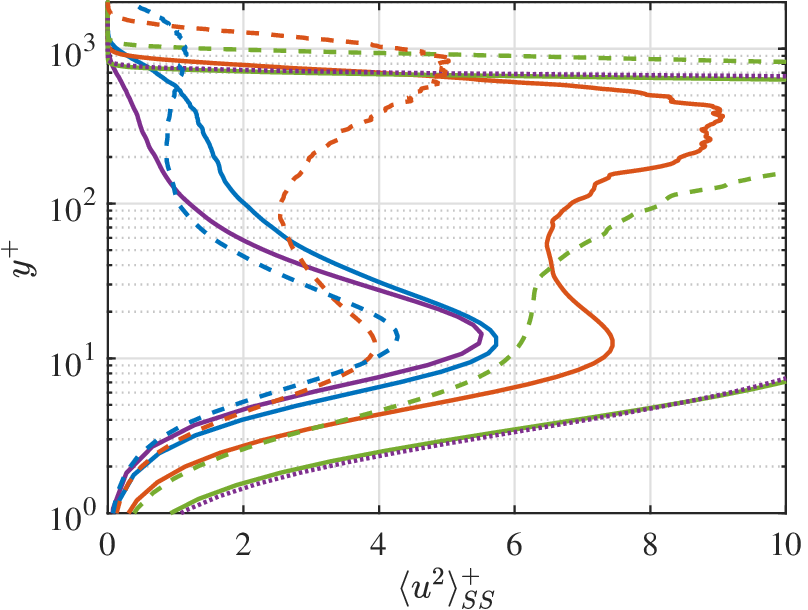}};
		\end{tikzpicture}  
		\hspace{0.05cm}			\begin{tikzpicture}   
			\node(a){ \includegraphics[scale=0.45]{ 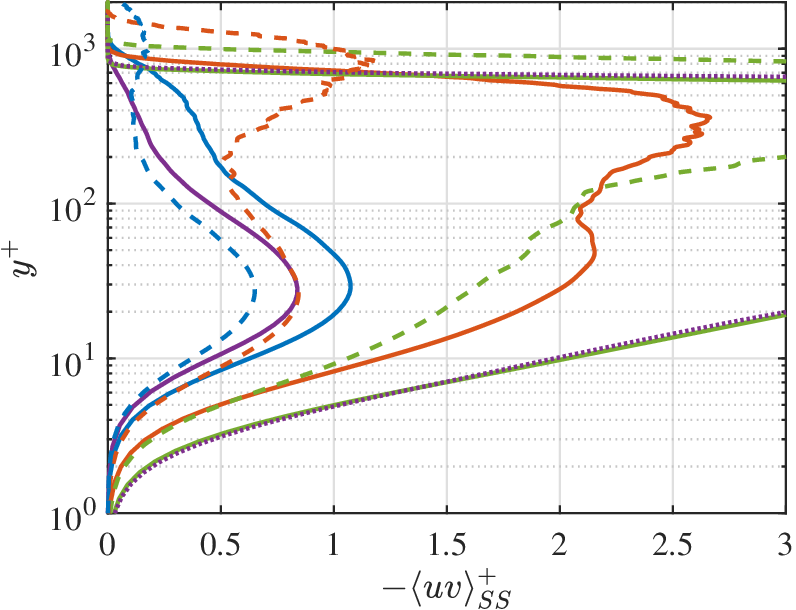}};
		\end{tikzpicture}

		\caption{The Reynolds stress profiles as a function of $y^+$ for the full signal (top row), large-scale structures (middle row), and small-scale structures (bottom row).}
		\label{fig:filtered}
	\end{figure}

	The spectral distributions of $\langle uv\rangle$ in figure \ref{fig:spectra_uv} exhibit similar results to those of $\langle u^2\rangle$. The inner peak is present in the small defect cases S2 and S8, although it is notably weaker at S8 when compared to the outer peak. The spectral distribution of $\langle uv\rangle$ at S2 is consistent with those found in the literature \citep{lee2017large,gungor2022energy}. In the large-defect case S5, the inner peak has vanished, and $\langle uv\rangle$ is significantly influenced by large-scale structures in the inner region, although small-scale structures still contribute. Overall, the effect of the large-scale structures in the inner layer is less important for $\langle uv\rangle$ than for $\langle u^2\rangle$, a trend also noted by \cite{gungor2024turbulent}.

	To investigate small-scale properties in the inner layer, we employ a filtering process to isolate the contributions of small scales by separating them from the large scales. To achieve this, we decompose the Reynolds stresses using a wavenumber sharp filter cutoff in the spectral domain, as depicted in figures \ref{fig:spectra_u2} and \ref{fig:spectra_uv} with vertical black dashed lines. We set the cutoff wavelength at $\lambda_z^+=300$, chosen for its effective differentiation between small and large scales in the inner region. This value is also used in the recent study by \cite{deshpande2023reynolds}. We have tested that the reconstructed Reynolds stress distributions are relatively insensitive, but only qualitatively (see discussion below), to the specific filter cutoff value.

	Figure \ref{fig:filtered} presents the Reynolds stress profiles of small scales ($\langle u^2 \rangle^+_{SS}$ and $\langle uv\rangle^+_{SS}$) and large scales ($\langle u^2 \rangle^+_{LS}$ and $\langle uv\rangle^+_{LS}$) as functions of $y^+$. The full Reynolds stress profiles are presented at the top of the figure to facilitate comparison. In contrast to the full $\langle u^2 \rangle^+$ profiles, the small-scale $\langle u^2 \rangle^+_{SS}$ profiles at S3 and S7 reveal an inner peak. Furthermore, the inner peak has become sharper at S1, S2 and S8. These results indicate that large-scale structures indeed influence the Reynolds stress distributions in the inner layer, leading to the obscuration of small-scale contributions and the near-wall cycle of turbulence production.

	\cite{deshpande2023reynolds} found that the inner peak of $\langle u^2 \rangle^+_{SS}$ collapses well across ZPG TBLs and TBLs subjected to very mild APG ($\beta_{RC}<1.7$), across various Reynolds numbers. The reported peak value of $\langle u^2 \rangle^+_{SS}$ is $5.5$ at $y^+=15$. Recall that we employ the same filtering procedure. In the current flow, the inner peak of $\langle u^2 \rangle^+_{SS}$ closely resembles that of \cite{deshpande2023reynolds} only at the first station, S1, with a small velocity defect and minimal history effects. As the velocity defect increases, the $\langle u^2 \rangle^+_{SS}$ inner profile also increases, eventually leading to the disappearance of the peak. Positions with substantial velocity defects, specifically S4 to S6, do not exhibit inner peaks in the profiles, except for a hump observed at S6. The disappearance of the near-wall peak suggests that the near-wall cycle could be significantly attenuated or even absent in TBLs with large velocity defects. 
	
	Nevertheless, interpreting the decomposed Reynolds stresses requires caution.  Firstly, the filtering procedure based on one-dimensional spectra has inherent limitations. A more refined separation between small and large scales could be achieved by employing a filtering approach based on two-dimensional streamwise-spanwise spectra, as demonstrated by \cite{lee2019spectral}. Similar to 	\cite{deshpande2023reynolds}, \cite{lee2019spectral} reported a Reynolds number independence of the $\langle u^2 \rangle^+_{SS}$ profile up to $y/\delta=0.2$ in channel flow cases. Yet, in their case, the peak value of $\langle u^2 \rangle^+_{SS}$ exceeded $7$. Lastly, it is plausible that the Reynolds number in our study might not be sufficiently high to achieve a clear-cut scale separation of scales.

	The inner peak of $\langle u^2 \rangle_{SS}^+$ at S7 is rather surprising because it is below that at S1, the latter closely following the ZPG TBL and small defect behavior presented in \cite{deshpande2023reynolds}. At S7, the pressure gradient is null. One could therefore expect the inner peak to be similar to the ZPG TBL one (and therefore also to the S1 peak) or even higher due to the upstream trend. The lower-level peak might suggest that the near-wall cycle has not yet fully recovered from the upstream pressure gradient effects, if it was indeed dampened or destroyed by them. It should be noted that, in contrast to the situation of $\langle u^2 \rangle_{SS}^+$, the inner peak of the small-scale Reynolds shear stress $-\langle uv \rangle_{SS}^+$ at S7 is similar to the one observed at S1. 
	
	Regarding the inner peak of $\langle u^2 \rangle_{SS}^+$ at S8, it is expected that it is lower than the one at S1. FPG TBLs that are in near equilibrium or are monotonically developing from a ZPG TBL have lower $\langle u^2 \rangle^+$ levels in the inner region (complete profile) compared to ZPG TBLs \citep{bourassa2009experimental,harun2013pressure,volino2020non}, even if the difference is not as pronounced as in the outer region.

	As for the small-scale Reynolds shear stress $-\langle uv \rangle_{SS}^+$, it behaves similarly to $\langle u^2 \rangle_{SS}^+$. The notable difference, as previously mentioned, is that the inner peak at S7 is similar to that at S1, as would be expected if there were no history effects. The inner peak of $-\langle uv \rangle_{SS}^+$ is associated with the near-wall production of $\langle u^2 \rangle_{SS}^+$. 
	
	The profiles of large-scale Reynolds stresses demonstrate that large scales indeed contribute significantly to the inner region, and this contribution increases with the mean velocity defect (not with $\beta_i$). As expected from the spectral distributions, $-\langle uv \rangle_{LS}^+$ is small below $y^+=10$ whereas $\langle u^2 \rangle_{LS}^+$ is significant in that region, comparable to $\langle u^2 \rangle_{SS}^+$. The large-scale structures reaching the very-near wall region cannot have a large wall-normal velocity. They are long streaky $u$-structures rather than sweeps and ejections.

	\begin{figure}
		\centering
		\vspace{0.5cm}
		\includegraphics[scale=0.6]{ 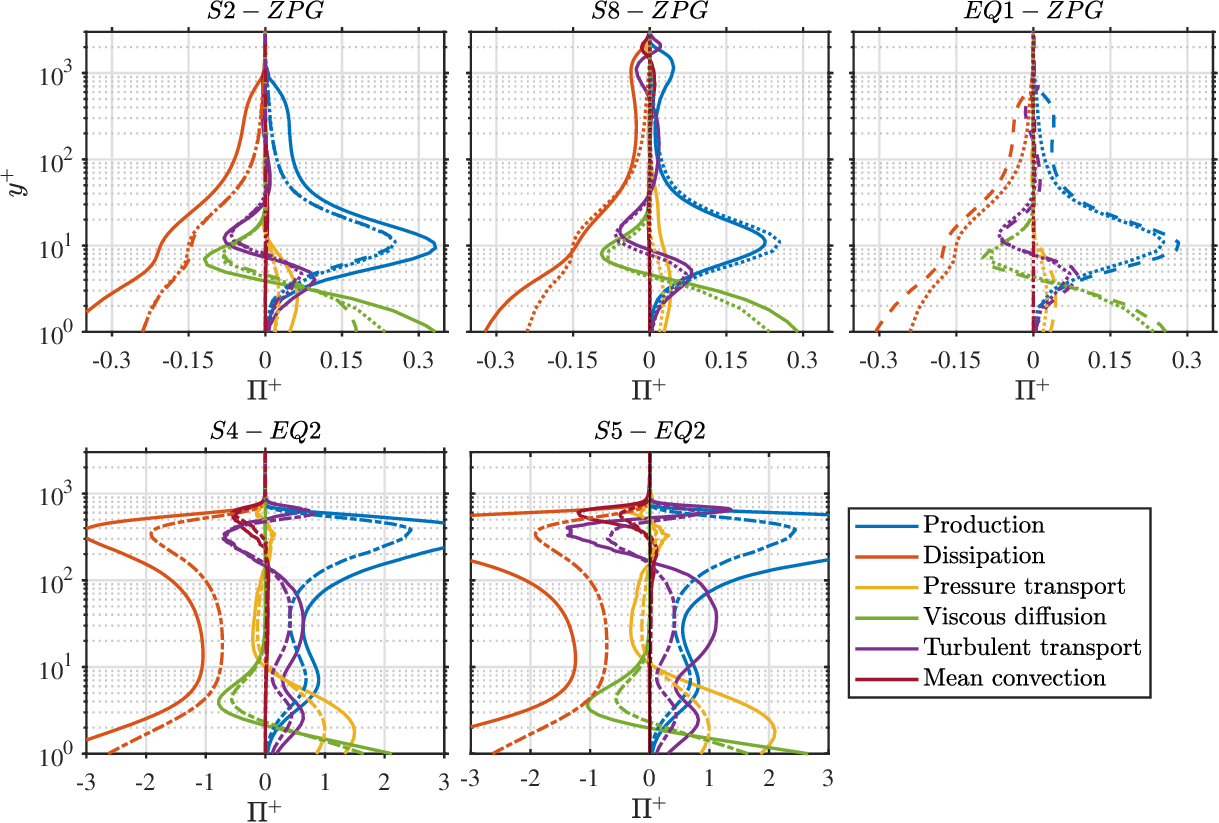} 	
		\caption{The turbulent kinetic energy budget for S2, S8, EQ1, ZPG, S4, S5 and EQ2 as a function of $y^+$. The terms are normalized with friction-viscous scales. The straight, dashed, dashed-dotted, and dotted lines are for the present case, EQ1, EQ2, and ZPG, respectively.}
		\label{inner_budget}
	\end{figure}
	
	To gain a deeper understanding of inner layer turbulence behavior, we analyze the turbulent kinetic energy budget at the streamwise positions S2, S8, S4, and S5 along with ZPG, EQ1, EQ2. Figure \ref{inner_budget} illustrates the budget terms, normalized with friction-viscous scales, as functions of $y^+$, with a similar format as the outer-scaled budgets in figure \ref{fig:outer_budget}. The balance of the budget terms in the inner layer at S2 closely resembles that of canonical wall flows; however, the levels of the terms are approximately $20$ to $30\%$ higher at S2. Another distinction is the higher level of production around $y^+=100$ relative to its peak value at $y^+=10$ in the APG case, aligning with the behavior of the Reynolds stresses depicted in figure \ref{fig:rs+}. These similarities and differences have been observed in small-defect APG TBLs, either originating as ZPG TBLs \citep{gungor2022energy}  or in near-equilibrium states \citep{kitsios2017direct,bobke2017history}
	
	Moving from S2 to S4 and S5, the inner wall-normal distributions of of the
	various energy transfer mechanisms become notably different. For instance, production still exhibits an inner maximum, but it is no longer dominant compared to the other terms due to high levels of turbulence being transported from the outer layer. The budget terms have higher values than those for EQ2, as can be expected given that $\beta_i$ is higher at S4 and S5. At S8, the inner budget has returned to one typical of canonical wall flows, with levels comparable to those of the ZPG TBL and smaller than at S2. However, turbulent transport remains strong above $y^+=40$ due to significant outer turbulence resulting from history effects. These history effects likely explain why the Reynolds normal stresses in the inner layer are much higher at S8 than at S2 (figure \ref{fig:rs+}) despite the lower energy transfer terms.

	\section{Conclusion}
	
	The current study investigates the response of the turbulent boundary layer (TBL) to uniform changes in the pressure force contribution to the force balance. The primary objective is to enhance our understanding and differentiation of the three types of effects resulting from the pressure force: the direct impact of the local pressure force, the influence of a local pressure force change (local disequilibrating effect), and the impact of the upstream pressure force evolution. The contribution of the pressure force to the force balance can be characterized by a pressure gradient parameter. However, due to differences in this contribution between the inner and outer regions (assuming two layers without consensus on this matter for pressure gradient boundary layers), two pressure gradient parameters become necessary. In this study, we have chosen the pressure gradient parameters $\beta_i$ and $\beta_{ZS}$ for the inner and outer regions, respectively, as we show that they follow reasonably well the ratio of pressure force to turbulent force in each region.
	
	To achieve the aforementioned objective, we devised a flow with a specific streamwise distribution of the pressure force contribution in the outer region, by choosing a distinctive distribution of $\beta_{ZS}$. The emphasis is placed on the outer region due to its more intricate and significant response compared to the near-wall region. Consequently, the pressure force contribution in the inner region becomes a result of this selection and, to a lesser extent, the Reynolds number of the flow. We designed and computed, using direct numerical simulation, a non-equilibrium flow characterized by a nearly linear increase followed by a nearly linear decrease of $\beta_{ZS}$. In the first zone, the uniformly increasing adverse-pressure-gradient (APG) impact leads to mean momentum loss in the boundary layer and an increase in turbulence when normalizing with $U_e$ or $u_{\tau}$. In the second zone, the constant $d\beta_{ZS}/dx<0$ results in two subzones: a first one of uniformly decreasing APG impact, and a final one of uniformly increasing favorable-pressure-gradient (FPG) impact. Despite this differentiation, the global effect in both subzones is the same: mean momentum gain in the boundary layer and a decrease in turbulence when normalizing with $U_e$ or $u_{\tau}$. Notably, in the first subzone with $d\beta_{ZS}/dx<0$, the flow exhibits a deliberately sought-after singular region where the boundary layer gains momentum while the external flow is still decelerating due to the adverse pressure gradient. 
	
	As anticipated, the mean flow's response to changes in the force balance differs significantly between the outer and inner layers. In the outer region, the results reveal a substantial delay in the mean flow response for both cases of momentum-losing ($d\beta_{ZS}/dx>0$) and momentum-gaining ($d\beta_{ZS}/dx<0$) pressure force effects. The order of magnitude of the delay is 10 boundary layer thicknesses. Comparisons with the two near-equilibrium APG TBLs of Kitsios et al. (2017) indicate that in the first momentum-losing zone, the delay results in an APG TBL with a smaller velocity defect compared to a near-equilibrium APG TBL at equivalent $\beta_{ZS}$ or $\beta_{RC}$. This is a manifestation of the delayed response to a local pressure force change. In the subsequent momentum-gaining zone, the prolonged delay leads to an FPG TBL at the end of the domain with a significant momentum defect in the outer region, comparable to that of APG TBLs with a moderate defect, and a hollowed-out mean velocity profile. These are clear effects of the upstream flow history, specifically the presence of a long momentum-losing zone upstream. 
	
	In the momentum-gaining zone, the increase in momentum initiates near the wall due to the rapid response of the inner region to pressure force changes. Very close to the wall, in the viscous sublayer, the mean flow responds almost instantaneously because inertia effects are negligible in that region. Inertia effects become noticeable above approximately $y^+=10$. Moving further away from the wall, the conventional logarithmic law of the ZPG TBL breaks down, even for cases with small defects in the current flow and in the near-equilibrium APG TBL of Kitsios et al. (2017). The local pressure force, therefore, has a direct impact on the logarithmic law. It is important to note that we cannot draw definitive conclusions about the logarithmic law due to the moderate Reynolds numbers of these three flows. However, the results of \cite{knopp2021experimental} at a Reynolds number an order of magnitude higher appear to confirm them. A positive $\beta_i$ results in a drop of $U^+$ below the log law and a change in the profile shape. This direct local effect of the pressure gradient can be accounted for in the initial part of the flow with the modifications of the logarithmic law proposed by \cite{nickels2004inner} and observed by \cite{knopp2021experimental}, using the local value of $\beta_i$. However, these modifications fail further downstream due to the non-equilibrium nature of the flow or the large values of $\beta_i$. The overlap layer is indeed greatly affected by upstream history and local changes in the pressure force. Streamwise locations with the same value of $\beta_i$ exhibit significantly different inner-scaled mean velocity profiles. One striking example occurs at a position where the pressure gradient is zero, yet the profile deviates considerably from that of the ZPG TBL.
	
	Turbulence responds with a delay to the mean flow changes in both the inner and outer regions, with the delay being much more prominent in the outer region. In the current flow, the pressure force contribution in the outer region was intentionally increased and then decreased with the same absolute amplitude ($|d\beta_{ZS}/dx|$). The outcomes indicate that outer turbulence takes significantly more time to decay in the second region than it takes to build up in the initial region. In fact, turbulence continues to intensify for a considerable length even after the pressure force begins to diminish, and its subsequent decay is very gradual. This slow decay of turbulence is not solely attributed to the persistence of intense turbulence structures convected from upstream. The results demonstrate that it also occurs because of the increase of the excess of turbulence production over its dissipation in the momentum-gaining zone of the flow. The Reynolds stress profiles also become fuller due to sustained high levels of turbulent transport. As a result of the slow decay of turbulence, the FPG TBL located at the end of the domain exhibits unusually high levels of outer turbulence, significantly surpassing those in the ZPG TBL.
	The comparison of Reynolds stresses in the outer region with those of the near-equilibrium APG TBLs by Kistios et al. (2017) also indicates a delay in the redistribution of energy form $\langle u^2 \rangle$ to $\langle v^2 \rangle$ and $\langle w^2 \rangle$ in the current flow. Furthermore, the outer maxima of all Reynolds stress components do not shift as far away from the wall as observed in the near-equilibrium cases. 
	
	Throughout the entire inner region, turbulence exhibits a delayed response to changes in the pressure force, even in the near-wall region below $y^+=10$, where mean inertia effects are negligible. Part of the reason for this delay can be attributed to the influence of large-scale turbulence on the near-wall region. The one-dimensional spanwise spectra of the Reynolds stresses reveal that large-scale structures with spanwise wavelengths on the order of $\delta$ significantly contribute to the Reynolds stresses in the inner region. This is particularly evident in the final momentum-gaining part of the flow, owing to the presence of energetic structures convected from upstream. Upon filtering out the large-scale contribution to the Reynolds stresses, the behavior of the small-scale Reynolds stresses aligns more consistently with the local value of $\beta_i$.  
	
	As documented in previous studies of both non-equilibrium and near-equilibrium large-defect APG TBLs, $\langle u^2 \rangle^+$ increases in the near-wall region as the local APG effect intensifies, similar to the behavior in the rest of the boundary layer. Simultaneously, the near-wall peak of $\langle u^2 \rangle^+$ becomes less sharp and eventually disappears in the large-velocity-defect zone of the flow. With the filtering out of the large-scale contributions, the inner peak of small-scale $\langle u^2 \rangle^+$ in the initial small-defect zone of the flow resembles that of the ZPG TBL. The peak is even present in large defect cases with $H$ around 2. However, it eventually disappears for larger defect cases, suggesting that the near-wall cycle could be significantly attenuated or even absent in TBLs with very large velocity defects.
	
	The current study contributes to our understanding of how the boundary layer responds to the three types of pressure force effects. Isolating the direct impact of the local pressure force from the other two effects is relatively straightforward. However, distinguishing between the local effect of a pressure force change (local disequilibrating effect) and the impact of the complete upstream pressure force history is challenging. This differentiation was achieved only in a few response situations in the present study. A more comprehensive distinction would necessitate various flow cases at high Reynolds numbers, encompassing a wide range of carefully selected pressure-gradient-parameter distributions. To complicate matters further, even if we provide convincing evidence for the appropriateness of the selected inner and outer pressure gradient parameters, this selection is still rightfully debatable.

	Not surprisingly, the significant cumulative effects of continuous pressure force variation suggest that characteristic parameters based only on local variables cannot fully capture the physics of non-equilibrium boundary layers. This limitation applies even to the streamwise derivative of the pressure gradient parameter ($d\beta/dx$), which characterizes the local disequilibrating effect of pressure force changes. Similarly, our preliminary investigation into existing response length scales, derived for localized step changes in pressure force, reveals their inadequacy in addressing continuous and gradual pressure force changes. Our research into finding appropriate parameters and response length scales for non-equilibrium TBLs is ongoing.

\acknowledgements{ \noindent \textbf{Acknowledgment.} We acknowledge PRACE for awarding us access to Marconi100 at CINECA, Italy and Calcul Québec (www.calculquebec.ca) and the Digital Research Alliance of Canada (alliancecan.ca) for awarding us access to Niagara HPC server. The authors would like to express their gratitude to Professors Javier Jiménez and Julio Soria for generously providing their data. 

\smallskip

\noindent \textbf{Funding.} TRG and YM acknowledge the support of the Natural Sciences and Engineering Research Council of Canada (NSERC), project number RGPIN-2019-04194. TRG and AGG were supported by the research funds of Istanbul Technical University (project numbers:MDK-2018-41689). 

\smallskip

\noindent \textbf{Declaration of interests.} The authors report no conflict of interest. 

\smallskip

\noindent \textbf{Data availability statement.} Statistical data for the current flow case are available at:  \\ \href{https://yvanmaciel.gmc.ulaval.ca/databases/}{https://yvanmaciel.gmc.ulaval.ca/databases/} and \href{https://web.itu.edu.tr/gungoray/databases/}{http://web.itu.edu.tr/gungoray/databases/} 

\smallskip

\noindent \textbf{AuthorORCID.} \\ T.R. Gungor, https://orcid.org/0000-0002-3143-8254; \\ A.G. Gungor, https://orcid.org/0000-
 0002-3501-9516; \\ Y. Maciel, https://orcid.org/0000-0003-1993-472X;}

	\appendix

	\section{The boundary layer thickness and edge velocity definitions}\label{appA}
	
	As discussed in the paper, defining the boundary layer thickness and edge velocity presents challenges for turbulent boundary layers subjected to a pressure gradient or over complex geometries where the freestream velocity varies. Here, we examine four methods outlined in the literature: the traditional $\delta_{99}$ method, a vorticity-based approach proposed by \cite{spalart1993experimental}, the method proposed by \cite{wei2023outer}, and the technique introduced by \cite{griffin2021general}. A comprehensive review of these methods, along with others, is available in \cite{griffin2021general}.
	
\begin{figure}
		\centering
		\vspace{0.5cm}
		\includegraphics[scale=0.4]{ 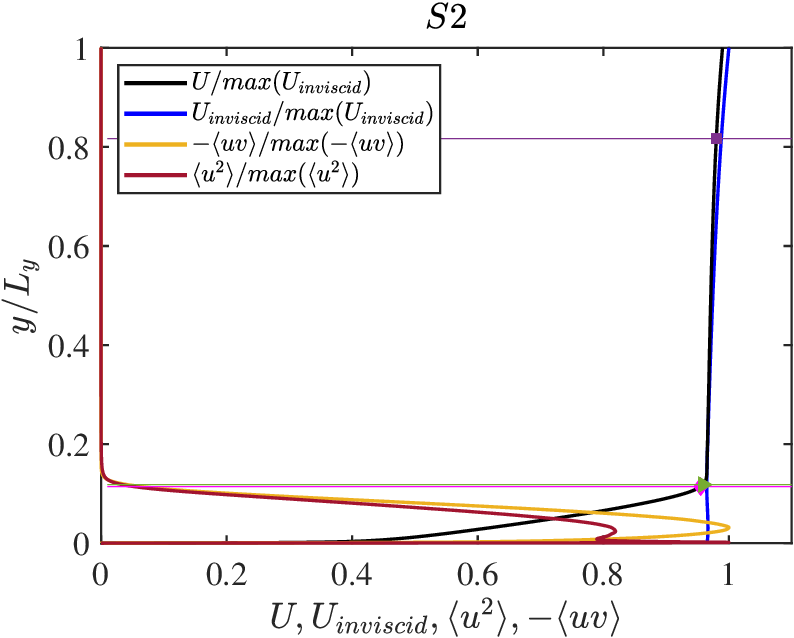} 	\includegraphics[scale=0.4]{ 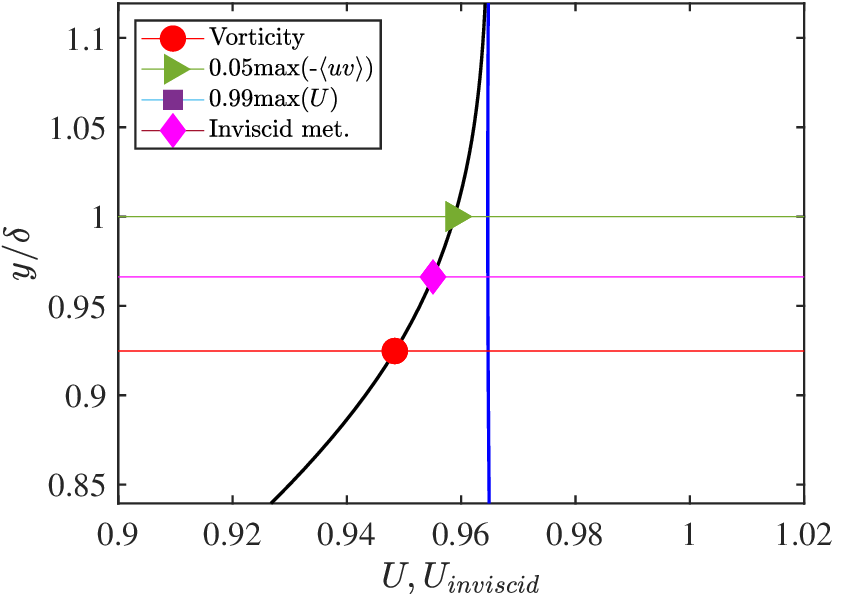} 
		
		\smallskip
		
		\includegraphics[scale=0.4]{ 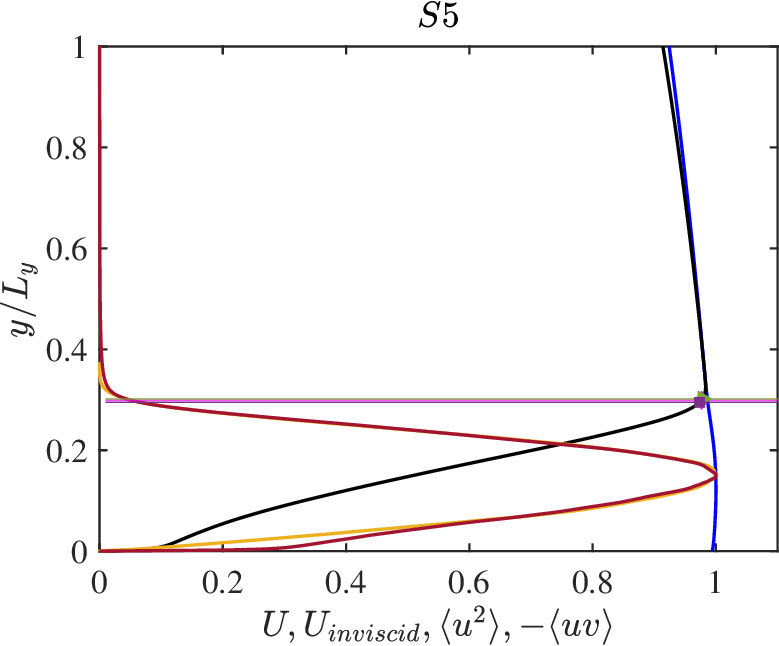} 	\includegraphics[scale=0.4]{ 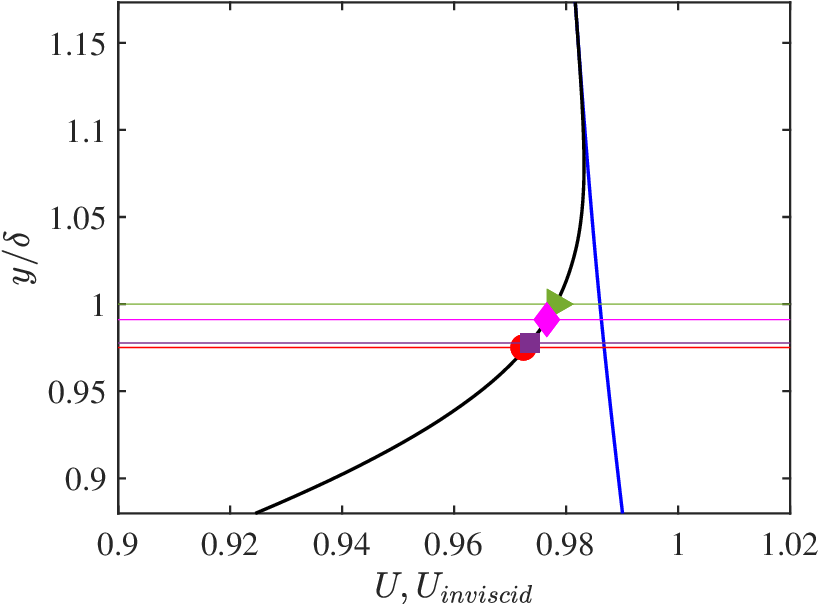} 
		
		\smallskip
		
		\includegraphics[scale=0.4]{ 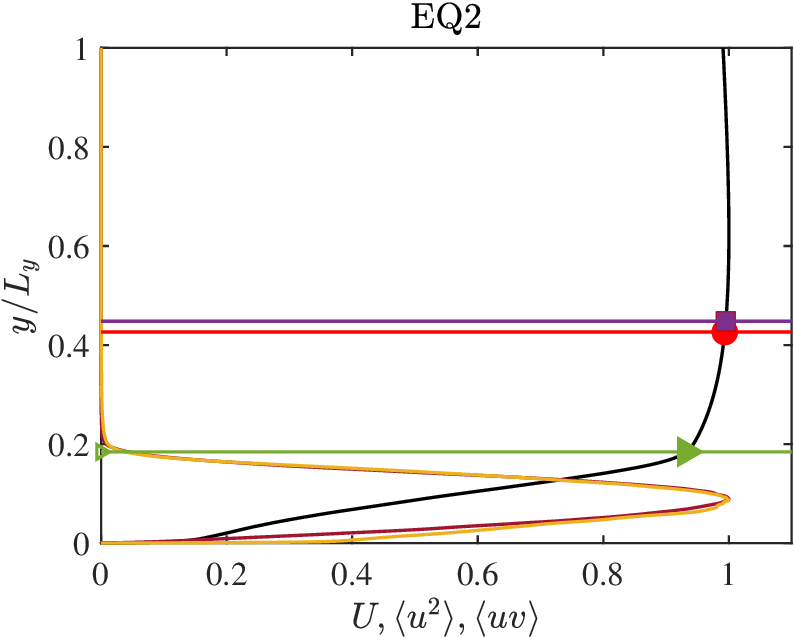} 		\includegraphics[scale=0.4]{ 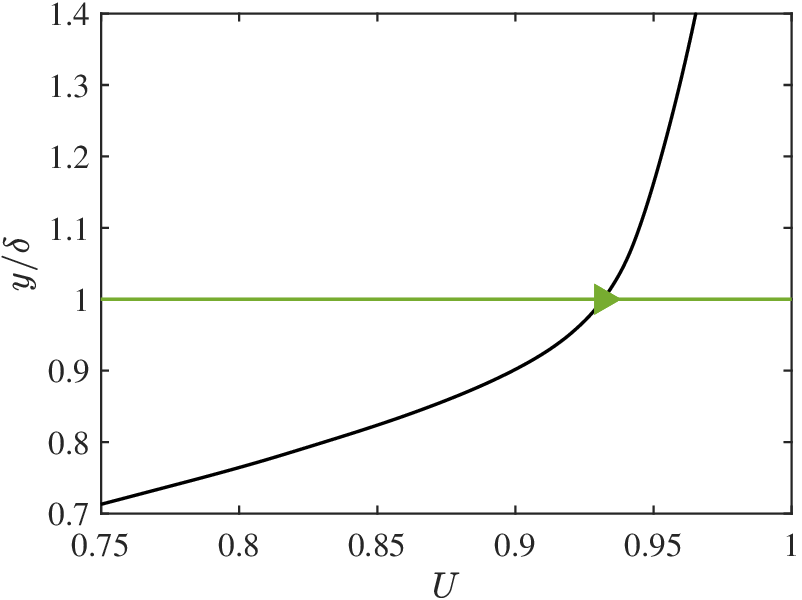} 			
		\caption{The mean velocity, inviscid velocity, $\langle uv \rangle$, and $\langle u^2\rangle$ profiles as a function of $y/L_y$ (left) and $y/\delta$ (right) where $\delta$ is computed using the boundary layer thickness definition in the paper. $L_y$ indicates the wall-normal height of the domain. }
		\label{app}
	\end{figure}

	Figure \ref{app} illustrates the comparison of the four methods for the present case and EQ2 of \cite{kitsios2017direct}. We do not present the results of the method by \cite{griffin2021general} for EQ2 because we do not have the mean pressure data. Nevertheless, the results indicate that the method is consistently reliable, aligning with the findings of \cite{griffin2021general}. It is also the most theoretically rigorous one. The method of \cite{wei2023outer} is also reliable. The results for the other two methods exhibit significant discrepancies. The traditional $\delta_{99}$ method, which is commonly employed in the literature, performs poorly at S2 and for EQ2. While the vorticity method yields satisfactory results for the present case, it does not work well for EQ2. It locates the boundary layer edge too high in the freestream. The method proposed by \cite{wei2023outer} remains consistent with the approach of \cite{griffin2021general} across all cases and works well for EQ2. Consequently, since we cannot use the method of \cite{griffin2021general} for EQ1 and EQ2 because we do not have the mean pressure data, we have opted for this method for the present study. The edge velocity is considered to be the velocity at the position of the boundary layer edge determined using the aforementioned method.

	\bibliography{references}
	\bibliographystyle{jfm}

\end{document}